\documentclass[lettersize,journal]{IEEEtran}
\usepackage{amsmath,amsfonts}
\usepackage{algorithmic}
\usepackage{algorithm}
\usepackage{array}
\usepackage[caption=false,font=normalsize,labelfont=sf,textfont=sf]{subfig}
\usepackage{textcomp}
\usepackage{stfloats}
\usepackage{url}
\usepackage{verbatim}
\usepackage{graphicx}
\usepackage{cite}
\usepackage{booktabs} 
\usepackage{etoolbox}
\makeatletter
\patchcmd{\@makecaption}
  {\scshape}
  {}
  {}
  {}
\makeatother
\begin{document}

\title{PuzzleTuning: Explicitly Bridge Pathological and Natural Image with Puzzles}

\author{Tianyi Zhang, Shangqing Lyu, Yanli Lei, Sicheng Chen, Nan Ying, Yufang He, Zeyu Liu, Yu Zhao, Yunlu Feng, Yueming Jin, Hwee Kuan Lee, Guanglei Zhang

\thanks{
Tianyi Zhang and Shangqing Lyu contributed equally to this work. Corresponding Author: Guanglei Zhang (e-mail: guangleizhang@buaa.edu.cn)

Tianyi Zhang, Yanli Lei, Nan Ying, Yufang He, Zeyu Liu and Guanglei Zhang are with the Beijing Advanced Innovation Center for Biomedical Engineering, School of Biological Science and Medical Engineering, Beihang University, Beijing, 100191, China (e-mails: \{zhangtianyi, 20373207, 20101014, heiheyhe, zeyuliu, guangleizhang\}@buaa.edu.cn)

Shangqing Lyu, Hwee Kuan Lee are with the Bioinformatics Institute (BII), Agency for Science, Technology and Research (A*STAR), Singapore 138410, Singapore (e-mail: shangqinglyu@outlook.com, leehk@bii.a-star.edu.sg)

Sicheng Chen is with the School of Microelectronics, Xi’an Jiaotong University, Xi’an 710049, China (e-mail: 2201111586@stu.xjtu.edu.cn)

Yu Zhao is with the Department of Pathology, Peking Union Medical College Hospital, Beijing 100006, China (e-mail: rain986532@126.com)

Yunlu Feng is with the Department of Gastroenterology, Peking Union Medical College Hospital, Beijing 100006, China (e-mail: yunluf@icloud.com)

Yueming Jin is with the Department of Biomedical Engineering, National University of Singapore, Singapore 117417, Singapore, and Department of Electrical and Computer Engineering, National University of Singapore, Singapore 117417, Singapore (e-mail: ymjin@nus.edu.sg)

Hwee Kuan Lee is aslo with the School of Computing, National University of Singapore, Singapore 117417, Singapore,
School of Biological Sciences, Nanyang Technological University, Singapore 639798, Singapore, and Singapore Eye Research Institute (SERI), Singapore 169856, Singapore, and International Research Laboratory on Artifical Intelligence, Singapore 138632, Singapore, and Singapore Institute for Clinical Sciences, Singapore 117609, Singapore (e-mail: leehk@bii.a-star.edu.sg)
}}



\maketitle

\begin{abstract}
Pathological image analysis is a crucial field in computer vision. Due to the annotation scarcity in the pathological field, pre-training with self-supervised learning (SSL) is widely applied to learn on unlabeled images. However, the current SSL-based pathological pre-training: (1) does not explicitly explore the essential focuses of the pathological field, and (2) does not effectively bridge with and thus take advantage of the knowledge from natural images. To explicitly address them, we propose our large-scale PuzzleTuning framework, containing the following innovations. Firstly, we define three task focuses that can effectively bridge knowledge of pathological and natural domain: appearance consistency, spatial consistency, and restoration understanding. Secondly, we devise a novel multiple puzzle restoring task, which explicitly pre-trains the model regarding these focuses. Thirdly, we introduce an explicit prompt-tuning process to incrementally integrate the domain-specific knowledge. It builds a bridge to align the large domain gap between natural and pathological images. Additionally, a curriculum-learning training strategy is designed to regulate task difficulty, making the model adaptive to the puzzle restoring complexity. Experimental results show that our PuzzleTuning framework outperforms the previous state-of-the-art methods in various downstream tasks on multiple datasets. The code, demo, and pre-trained weights are available at https://github.com/sagizty/PuzzleTuning.
\end{abstract}

\section{Introduction}
\label{sec:intro}

\IEEEPARstart{P}athological image analysis is the gold standard in various medical diagnoses \cite{transpath, ying2023cpia}. Traditional pathological examination depends on the judgment of experts under the microscope, which is inefficient and may be affected by factors such as experience or fatigue \cite{transpath, ying2023cpia, models_genesis}. In recent years, the trending Deep learning (DL) approaches has been studied to automate the pathological diagnosis \cite{ying2023cpia, models_genesis, nsclc_evo, domain_adapt_mia}. 

However, the performance of DL-based pathological image analysis highly relies on the model initialization \cite{ying2023cpia, models_genesis, nsclc_evo, domain_adapt_mia}. Accordingly, the pre-training is proposed to improve the initialization of models with large-scale data before finetuning them on specific downstream tasks. However, the labeled pathological images are of great scarcity which set limitations for initializing models through supervised pre-training. Studies \cite{transpath, ying2023cpia, sgcl, chitnis2023domainspecific} relieve this challenge with the trending self-supervised learning (SSL) to pre-train models. Meanwhile, most researchers directly apply transfer learning \cite{ivit, lu2021ai, gtp, clam, CellViT}. Via initializing weights after pre-training on large-scale natural datasets such as ImageNet \cite{deng2009imagenet}, models are significantly benefited from general vision knowledge from natural images \cite{ying2023cpia, zhang2022shuffle, zhang2023cellmix}. 

Still, the inevitable domain gap between the natural and pathological images hinders the eventual performance of models with current SSL pre-training and transfer learning \cite{ying2023cpia, domain_adapt_mia}. Specifically, compared with natural images, the pathological images present local similarities and global heterogeneity \cite{zhang2022shuffle, chitnis2023domainspecific, gcmae}. Additionally, tissues exhibit different biological entities at different length scales, which makes their features discretely distributed in various scales \cite{transpath, ying2023cpia, sgcl}. To improve the model initialization, we aim to take advantage of general vision knowledge and encoding it with pathological focuses, where large-scale SSL pre-training is expected to enhance downstream performance and reliability. Correspondingly, two key challenges are identified for improving SSL pre-training in pathological image analysis: (1) The pre-training task should have explicit focuses on pathological characteristics. (2) The general vision knowledge should be effectively seamed with pathological knowledge to improve the generalization abilities on downstream pathological tasks.

\begin{figure*}
  \centering
    \includegraphics[width=1.0\linewidth]{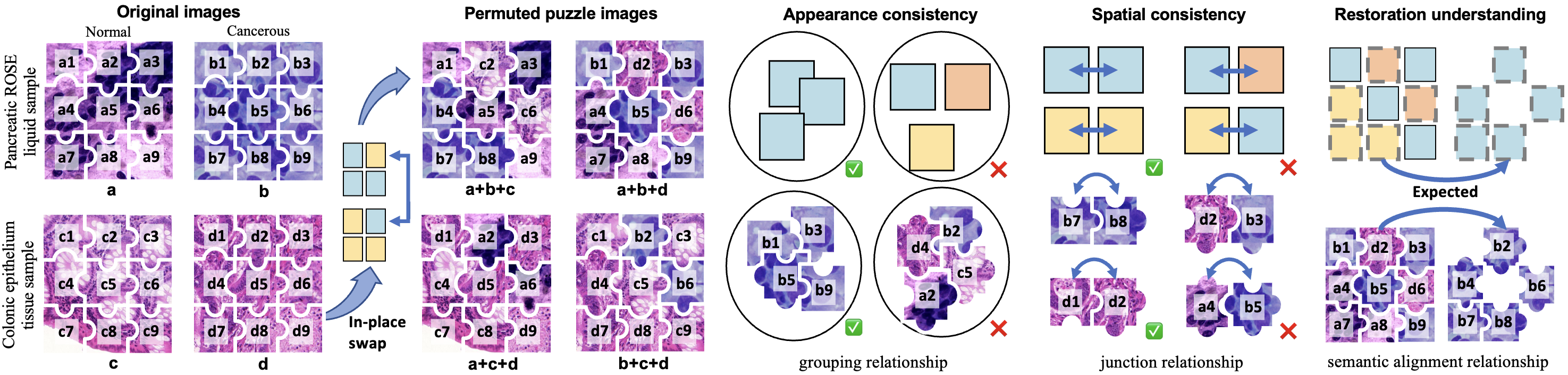}
  \caption{Samples illustrate the focuses and relationships in pathological images. They are pancreatic liquid samples (a and b) and colonic epithelium tissue samples (c and d) of normal (a and c) and cancer conditions (b and d). The patches of them are numbered from 1 to 9. Grouping the patches from each image as a bag, after intermixing patches among them, the three pathological focuses of appearance consistency, spatial consistency, and restoration understanding are highlighted.}
  \label{fig:fig_concept}
\end{figure*}

To learn pathological characteristics explicitly in pre-training, the patch-based modeling in \cite{zhang2023cellmix, zhang2022shuffle} and the jigsaw puzzle task in \cite{jigsaw} inspire us to focus on identifying relationship. We define the relational focuses, which are the task objectives of learning the relationship within the local neighbourhood of pixels (patch). Specifically, Fig. \ref{fig:fig_concept} is illustrated by cropping the image into patches (numbered 1-9), and regrouping the patches of pancreatic liquid samples (a and b) and colonic epithelium tissue samples (c and d) of normal and cancerous conditions. Accordingly, we identified three key task focuses: appearance consistency, spatial consistency, and restoration understanding.

{\bf Appearance consistency}. Pathological images exhibit local homogeneity. In Fig. \ref{fig:fig_concept}, when grouping the patches from one image as a bag, the patches from the same image consistently present similar appearance (patches within groups a, b, c, and d). In contrast, when grouping bags with intermixed patches from different organs (a+c, b+d) or different conditions within the same organ (mixing cancerous and normal patches with a+b, c+d), the co-occurrence of patches in these bags highlights heterogeneity. The consistent/conflicting features in the bag \cite{zhang2022shuffle} is defined as grouping relationships. Explicitly focusing on such in the learning process \cite{transpath} can enhance the identification of different pathological features \cite{zhang2023cellmix}.

{\bf Spatial consistency}. Adjacent patches in pathological images inherently share consistent spatial relations. In Fig. \ref{fig:fig_concept}, when grouping the patches from one image as a bag, all patches numbered 4,7,8 (from grouped bags a, b, c, and d) are fixed while others are shuffled to other bags. After shuffling, the consistency of pixel level textures are preserved when the adjacent patches come from the same bag. Meanwhile, the patches from different bags present texture misalignment. Accordingly, we define the spatial consistency among the adjacent patches as junction relationship. Extrapolating such junction relationships can help models restore accurate tissue and cell cluster compositions \cite{mae, simmim, Kong_2023_CVPR, luo2023selfdistillation}.

{\bf Restoration understanding}. Opposite to understanding the appearance and spatial consistencies, correcting the misalignment scenarios offers another effective contrast learning goal \cite{transpath, jigsaw}. Specifically, learning the restoration among multiple swapped grouped bags (1,3,5,9) vs. (4,7,8) explicitly focuses on understanding the semantic alignment relationships. Additionally, pathological images present various feature scales \cite{zhang2023cellmix}, ranging from individual cells (each patch in b), cell clusters (group of (b4,b7,b8) and (a4,a7,a8)), gland to tissue structures (c and d). Through restoring the patches based on their feature representation, the multi-scale understanding of feature alignment can be explicitly learned. 

However, regarding the first challenge, current SSL pre-training task designs for pathological images \cite{virchow, gcmae, transpath} does not explicitly pinpoint the three key task focuses of appearance consistency, spatial consistency, and restoration understanding. Specifically, the current state-of-the-art (SOTA) SSL paradigm can be grouped into Masked Image Modeling (MIM) \cite{mae, gcmae, BeyondMask, wang2023droppos,  simmim} and Contrastive Learning (CL) \cite{virchow, jigsaw, moco, simclr, byol}. 
Firstly, MIMs learn occlusion invariant feature representations through reconstructing \cite{mae, gcmae, Kong_2023_CVPR}, which effectively captures spatial semantics on adjacent patches. However, it struggles with the homogeneity characteristic of pathological images, which often exhibit repetitive spatial features. Therefore, beyond identifying junction relationships in natural images, advanced pathological modeling demands explicit learning on the feature distribution, emphasizing the groupings and semantic alignment relationships.
Then, while CLs identify grouping relationships based on general representations \cite{sgcl, moco, simclr, byol}, their training effectiveness is highly constrained due to the limited variety of categories and narrower knowledge scope in pathological images \cite{transpath, sgcl}.

Moreover, regarding the second challenge, general vision knowledge should be further exploited through effective adaptations between natural and pathological images. Firstly, transferring the general vision knowledge brings significant improvements \cite{transpath, ying2023cpia, zhang2023cellmix, zhang2022shuffle, sgcl, chitnis2023domainspecific}. However, pathological datasets are generally smaller and have a narrower knowledge scope than the diverse patterns in natural image datasets \cite{ying2023cpia, domain_adapt_mia, chitnis2023domainspecific}. Accordingly, several previous pathology pre-training attempts \cite{transpath, virchow, gcmae, luo2023selfdistillation, emerge_ssl_vit} neglect general knowledge from natural images to reduce the domain gap. 
Recently, to take advantage of both general and domain-specific knowledge \cite{domain_adapt_mia}, domain-bridging learning has been explored. However, most approaches finetune all parameters and move them from natural to pathological domain \cite{ying2023cpia, zhang2022shuffle, virchow, gcmae}. Consequently, this implicit knowledge bridging overwrites the original general vision knowledge, which limits its generalization potential \cite{domain_adapt_mia}.

To address the challenges explicitly, we aim to bridge the pathological and natural images with an explicit SSL pre-training task and additional bridging parameters. First, we designed a novel SSL task to restore multiple puzzles for an enormous scale of unlabeled pathological images \cite{ying2023cpia}. Accordingly, this SSL task explicitly focuses on the grouping, junction, and semantic alignment relationships at multiple scales. Specifically, a batch of input pathological images is broken into patches, and a fraction of these patches are shuffled among different images, tuning the batch into a state called ‘Puzzle’. Then, with a hint of the starting patches, the model is trained to restore the original image batch. Through the training, the model understands the grouping within multiple images, as well as the spatial junction relationships among the patches. By traversing multiple patch scales, the shuffling process introduces the multi-scale semantic alignment relationship. Therefore, multiple puzzle restoring task can effectively teach appearance consistency, spatial consistency, and restoration understanding.

Then, we explicitly bridge pathological and natural general vision knowledge by pre-training the models (trained on ImageNet \cite{deng2009imagenet}) on pathological images, before finetuning them on the downstream-specific tasks. Moreover, we designed a prompting-based dataflow with additional prompt tokens \cite{vpt}. Specifically, we begin by initializing our model with weights pre-trained on natural images, encapsulating general vision knowledge. While fixing backbone parameters, we then update the prompt tokens using pathological images during training. Lastly, after pre-training, we finetune all parameters on downstream tasks. Additionally, we apply curriculum learning \cite{curriculum_learning} to bridge the domain divergence between natural and pathological fields adaptively. This strategy emulates the human easy-to-hard learning trajectory, a method verified for its efficacy in complex tasks \cite{zhang2023cellmix, curriculum_learning}. Our practice is gauged by puzzle shuffling intricacy, starting with simpler mixing and increasing shuffling complexity over training. Covering multiple feature scales, we initiate pre-training with larger puzzle patches, narrowing down over epochs. This propagation in complexity and granularity bolsters the capacity for semantic understanding, adapting to the semantic gap of different images. With this strategy, our SSL pre-training approach is termed 'PuzzleTuning'.

In summary, our contributions are of four aspects:

1. Task focuses for pathology: By observing grouping, junction, and semantic alignment relationships at multiple scales, we define three crucial task focuses in pathological pre-training: appearance consistency, spatial consistency, and restoration understanding. Based on them, we present a novel SSL pre-training task of multiple puzzles restoring, termed as 'PuzzleTuning'.

2. Domain bridging with prompt-tuning: To effectively take advantage of general vision knowledge, we employ an explicit prompt-tuning technique. The prompt tokens, attached to the backbone, explicitly learn the bridging knowledge that seams the two domains.

3. Curriculum learning: We introduce an innovative curriculum learning strategy that modulates the PuzzleTuning training difficulties, seamlessly adapting the multi-scale pathological focuses and general knowledge.

4. Large-scale SSL in pathology: Incorporating over 100 datasets of wide scope, this research is one of the most expansive SSL in pathological image analysis. Experimental results indicate that PuzzleTuning has surpassed previous SOTA methods in multiple downstream tasks.

\section{Related Works}
\label{sec:related_works}

SSL pre-training can exploit large unannotated pathological image datasets for model initialization. Specifically, CL approaches are grounded in the premise that visually similar objects consistently possess identical labels. Widely applied CL representation learning frameworks, such as Virchow \cite{virchow} based on DINO \cite{emerge_ssl_vit}, TransPath \cite{transpath} based on BYOL \cite{byol}, MoCo \cite{moco}, and SimCLR \cite{simclr} based on Siamese architectures, have become staple frameworks in pathological image pre-training. Additionally, the CL-based pre-training is also explored with task-specific objectives such as semantic segmentation in medical image analysis \cite{haq2022self, chen2023contrastive}. Their training aims to learn representations that amplify alignment between analogous scenes while concurrently reducing similarities among unrelated images \cite{moco, simclr, byol}.

Another SSL paradigm, MIM, is gaining prevalence over CL in pathological image analysis, attributed to its superior performance and training efficacy. The most representative MIM, Masked Auto Encoder (MAE) \cite{mae}, randomly masks a large proportion of input images and reconstructs the obscured pixels through a lightweight decoder. Recent studies spotlight advancements by advancing occlusion invariant learning \cite{Kong_2023_CVPR} in MAE.

Firstly, focusing on the modeling process in MIM, CAE \cite{CAE} builds the additional module that explicitly separates the low-level reconstruction task aiming for better abstractive focuses in the backbone training. However, recent studies that emerged in pathological image analysis highlight the modeling process of MAE, such as GC-MAE \cite{gcmae} integrate local feature extraction of MAE with the global focus of contrast learning.
Focusing on feature representation, MaskFeat \cite{MaskFeat} proposes patch-free modeling objectives instead of masking, while approaches such as DropPos \cite{wang2023droppos} emphasize the visible patches in MAE. Inspired by the early work of Jigsaw \cite{jigsaw}, those methods utilize a sequence-restoring mechanism to refine the junction modeling. Rather than extrapolating such junction relationship in sequence, the Siamese structured methods SimMIM \cite{simmim} and BeyondMask \cite{BeyondMask} are proposed with multiple pretext objectives that integrate the CL and MIM learning. Nevertheless, the current SSL has not explicitly explore the highlighted grouping, junction and semantic alignment relationships in pathological images. 

Moreover, the existing domain gap between natural and pathological images calls for effective alignment. Prompt tuning, proposed to bridge knowledge gaps, has become a prevalent method in pre-trained models of language and vision fields \cite{vpt, li2021prefixtuning}. In prompt tuning, the parameters of the original pre-trained model are fixed, while only some additional parameters acting as prompt tokens are updated during pre-training. Serving as an incremental learning procedure \cite{li2021prefixtuning, lester2021power}, prompt tokens explicitly carry extra information to align downstream tasks. Rather than finetuning and resetting parameters with implicit learning in a new domain, explicit prompt-based structures can effectively encode additional domain bridging knowledge. However, there are few existing works in pathological image analysis to leverage this trending approach for bridging knowledge domains during pre-training.
\section{Methods}
\label{sec:methods}

\begin{figure*}
  \centering
    \includegraphics[width=0.9\linewidth]{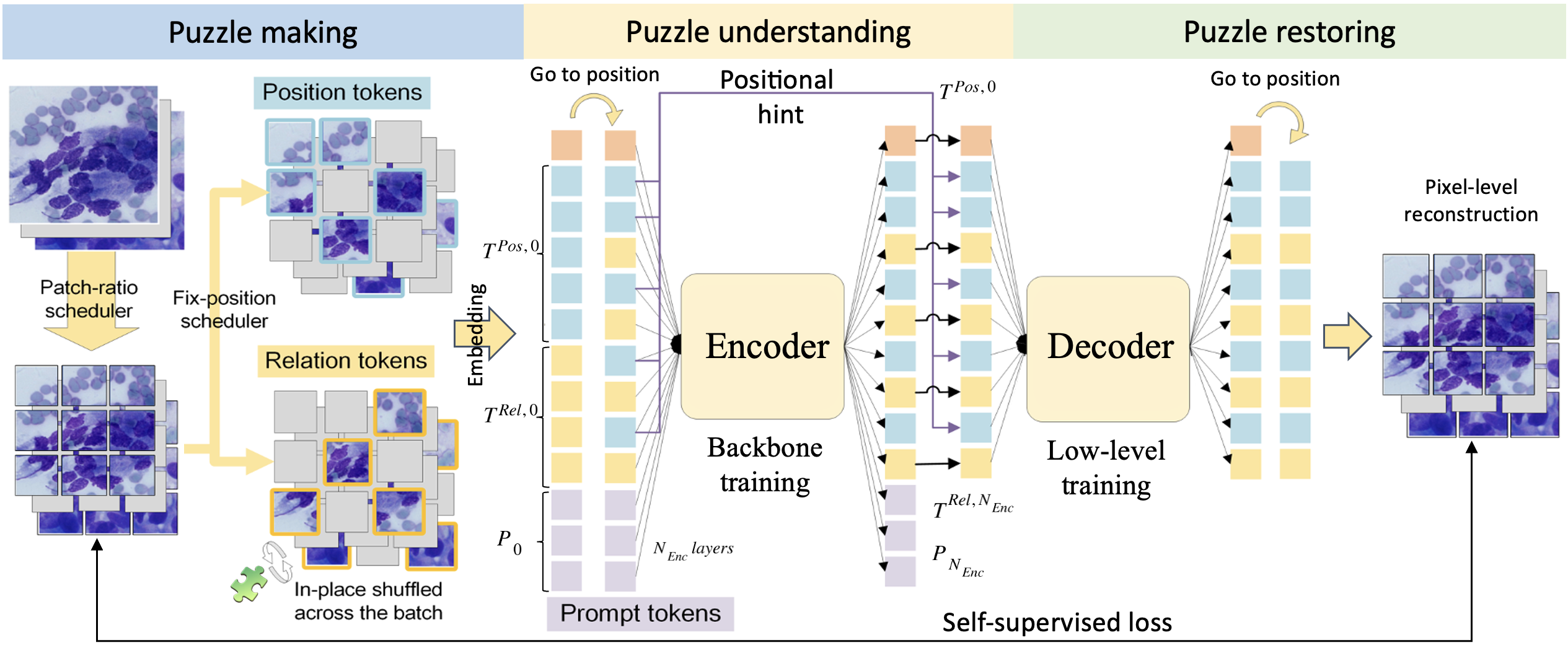}
  \caption{Three steps are designed in PuzzleTuning: 1) Puzzle making, where image batch are divided into bags of patches and fix-position and relation identity are randomly assigned. The relation patches are then in-place shuffled with each other, making up the puzzle state. 2) Puzzle understanding, where puzzles regarding grouping, junction, and restoration relationships are learned by prompt tokens attached to the encoder. Through the prompt tokens, the pathological focuses are explicitly seamed with general vision knowledge. 3) Puzzle restoring, where the decoder restores the relation patches with position patches as hint, under SSL supervision against original images.}
  
  \label{fig:fig_PuzzleTuning_method}
\end{figure*}

As illustrated in Fig. \ref{fig:fig_PuzzleTuning_method}, three main processes are designed to explicitly bridge the domains with pathological focuses, including (1) Puzzle making process, (2) Puzzle understanding process, and (3) Puzzle restoring process.

\subsection{Puzzle Making Process}

In the puzzle making process, we primarily introduce the shuffle strategy formulating the multiple puzzles, which explicitly train the model to mine the grouping, junction, and semantic alignment relationships with pathological images. Next, we traverse different sizes of the patch to explore the multi-scale relationship. From easy to hard with curriculum learning, we modulate puzzle mixing complexity.

\subsubsection{Multi-sample puzzle making}

\begin{figure}[t]
  \centering
   \includegraphics[width=1.0\linewidth]{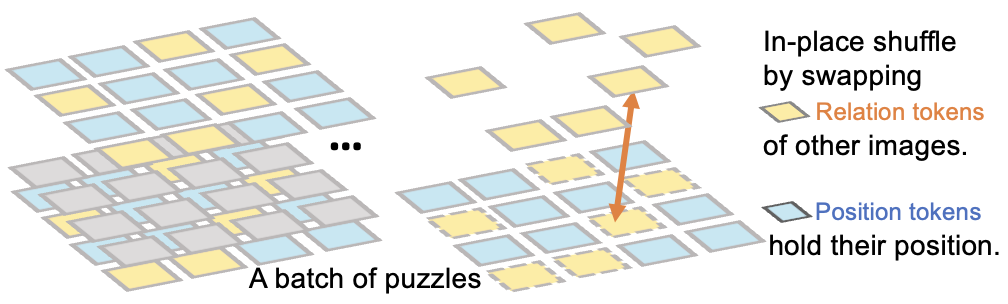}
   \caption{The in-place shuffle: the relation patches across a group of different images are swapped with each other.}
   \label{fig:fig_shuffle}
\end{figure}

In puzzle making at Fig. \ref{fig:fig_PuzzleTuning_method}, we have a batch of $B$ pathological images ($\tilde{I}_b$ with a size of $3,h,w$ pixels, where $b = 1, \cdots, B$). Each $\tilde{I}_b$ is cropped into a set of $m$ patches (size of $(3, P, P)$ pixels), which forms up a bag $I_b=\{I_{1, b}, \cdots, I_{m, b}\}$. Next, following the location index $x \in \{1, \cdots, m\}$ of the patches, we randomly assign $r$ out of $m$ positions under a given fix-position ratio $F_r$. 

As shown in Fig. \ref{fig:fig_shuffle}, the chosen $r$ patches are designed as position patches shown in blue, while the rest of $m-r$ patches are designed as relation patches in yellow. Then, we apply the same selection to all the $B$ bags. Accordingly, each bag $I_b$ is split into two disjoint subsets $I_{b}^{Pos}$ for $r$ position patches designed to hold their positions, and $I_{b}^{Rel}$ for the $m-r$ relation patches to be shuffled. This process generates each bag $I_b = \{I_{b}^{Pos} \cup I_{b}^{Rel}\}$ as 

\begin{equation}
    I_b = \{\{I_{1,b}^{Pos}, \cdots, I_{r,b}^{Pos}\} \cup \{I_{1,b}^{Rel}, \cdots, I_{m-r,b}^{Rel}\}\}
  \label{eq:puzzle_making_2_1}
\end{equation}

Then, as illustrated in Fig. \ref{fig:fig_shuffle}, the relation patches $I_{b}^{Rel}$ from puzzle batch $B$ are randomly swapped while position patches $I_{b}^{Pos}$ all stay unchanged. This in-place shuffle is done across bags while keeping the $(X,Y)$ coordinates of the patches fixed. It re-assigns the batch indexes $b_{i}$ of the $i$-th relation patches, accordingly, $I_{b}^{Rel}$ is changed to $\hat{I}_{b}^{Rel}$ where

\begin{equation}
    \hat{I}_{b}^{Rel} = \left\{I_{1,b_{1}}^{Rel}, I_{2,b_{2}}^{Rel}, \dots, I_{m-r,b_{m-r}}^{Rel} \big|b_{i} \in \{1, \dots, B\} \right\}
  \label{eq:puzzle_making_2_2}
\end{equation}

Accordingly, this step makes up a puzzle state $B_{puzzle}=\{\hat{I}_b|b = 1, \dots, B\}$ where each permuted bag $\hat{I}_b$ is defined as $\hat{I}_b = \{I_{b}^{Pos} \cup \hat{I}_{b}^{Rel}\}$. 
After shuffling, among the patches from the same original bag $I_b$, their absolute relationship remains unchanged. Meanwhile, among permuted $\hat{I}_{b}^{Rel}$ and fixed $I^{Pos}_{b}$ patches, the design changes grouping and semantic alignment relationships in bag $\hat{I}_b$. As shuffling creates new junction relationships of the patches, a series of puzzles presenting the three pathological focuses are ready to be learned. 

\subsubsection{Curriculum design of puzzles}

\begin{figure}[t]
  \centering
   \includegraphics[width=1.0\linewidth]{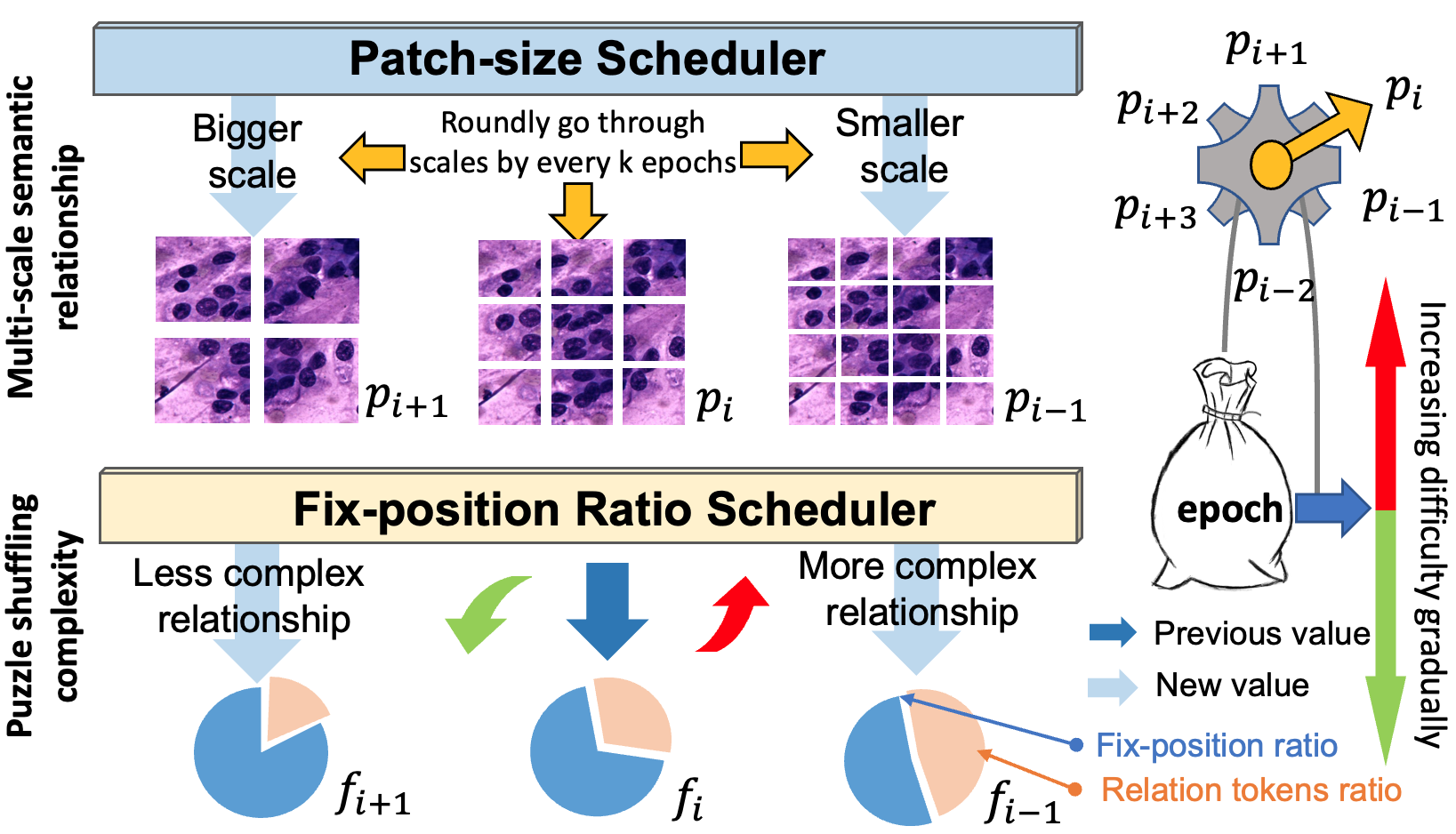}
   \caption{The curriculum learning design regulates puzzle complexity therefore altering difficulty. During training, the difficulty gradually increases by decreasing the fix-position ratio. Additionally, the patch size roundly goes through all scales to capture the relationship at multiple scales.}
   \label{fig:fig_curriculum_learning}
\end{figure}

As shown in Fig. \ref{fig:fig_curriculum_learning}, with the curriculum learning concept, we build puzzles from easy to hard. During training, we gradually reduce the fix-position ratio $F_r=f_e$ with a fix-position ratio scheduler on epoch $e$. Reducing $f_e$ from 90\% to 20\%, the shuffling complexity of permuted $I_{b}^{Rel}$ is progressively increased. Accordingly, the task complexity of learning the patch relationship is gradually increased \cite{zhang2023cellmix}.

Considering the multi-scale semantic alignment relationships, the granularity varies with the scale of pathological features. Controlled by a patch-size scheduler shown in Fig. \ref{fig:fig_curriculum_learning}, on epoch $e$, we traverse the patch size $P=p_e$ to generate each bag $I_b$ with a set of $m$ patches, size of $(3, P, P)$ pixels. Specifically, in Fig. \ref{fig:fig_curriculum_learning}, we employ a repetitive-looping strategy, cycling patch sizes from 16 to 112 every three epochs. It enables a more generalized and robust model training with multiple feature scales of patch $I_{x,b}^{Pos}$ and $I_{x,b}^{Rel}$. 

\subsection{Puzzle Understanding Process}

We explicitly seam the pathological focuses with general vision understandings in the puzzle understanding process. Firstly, the $N_{Enc}$ layer encoder ViT \cite{vit} is built to model the identifications and relationships of the patches \cite{zhang2022shuffle}. Then, based on Visual Prompt Tuning (VPT) \cite{vpt}, we introduce additional prompt tokens into it, bridging the domains explicitly. Lastly, we design a special positional hint in the dataflow, which enables the proper restoration of shuffled puzzle bags.

\subsubsection{Encoder backbone}

The ViT is applied to model the abstract relationships of the puzzle patches. For the ViT with $N_{Enc}$ encoder layers $L_{n}, n=1,\cdots,N_{Enc}$, there are $k_{Enc} = m +1$ tokens of $D_{Enc}$ dimension are learnt. Specifically, each bag $\hat{I}_b$ of the input puzzle batch $B_{puzzle}$ has its position and relation patches $I_{x,b}^{Pos}$ and $I_{x,b}^{Rel}$ embedded as position tokens $T_{x,b}^{Pos,0}$ and relation tokens $T_{x,b}^{Rel,0}$. Accordingly, we denote the embedded tokens as $T_{b}^{Pos,0} =\{ T_{x,b}^{Pos,0} | x = 1, \dots, r\}$ and $T_{b}^{Rel,0} =\{ T_{x,b}^{Rel,0} | x = 1, \dots, m-r\}$.

Following the ViT \cite{vit}, the learnable positional embedding is added to each token and then an additional classification token $T_{b}^{cls,0}$ is concatenated. The input bag $\hat{I}_b$ for puzzle understanding becomes $T_{b}^{Enc,0}=[T_{b}^{cls,0}, T_{b}^{Pos,0}, T_{b}^{Rel,0}]$. Then, the $N_{Enc}$ layers of relationship encoding are formulated as

\begin{equation}
  \resizebox{\linewidth}{!}{
    $[T_{b}^{cls,n},T_{b}^{Pos,n},T_{b}^{Rel,n}] = L_n\left[T_{b}^{cls,n-1},T_{b}^{Pos,n-1},T_{b}^{Rel,n-1} \right]$}
  \label{eq:encoder_backbone_2}
\end{equation}

After encoder modeling, all the output tokens $[T_{b}^{cls,N_{Enc}}, T_{b}^{Pos,N_{Enc}}, T_{b}^{Rel,N_{Enc}}]$ of the last layer $N_{Enc}$ are used.

\subsubsection{Prompt tokens for domain bridging}

\begin{figure}[t]
  \centering
   \includegraphics[width=1.0\linewidth]{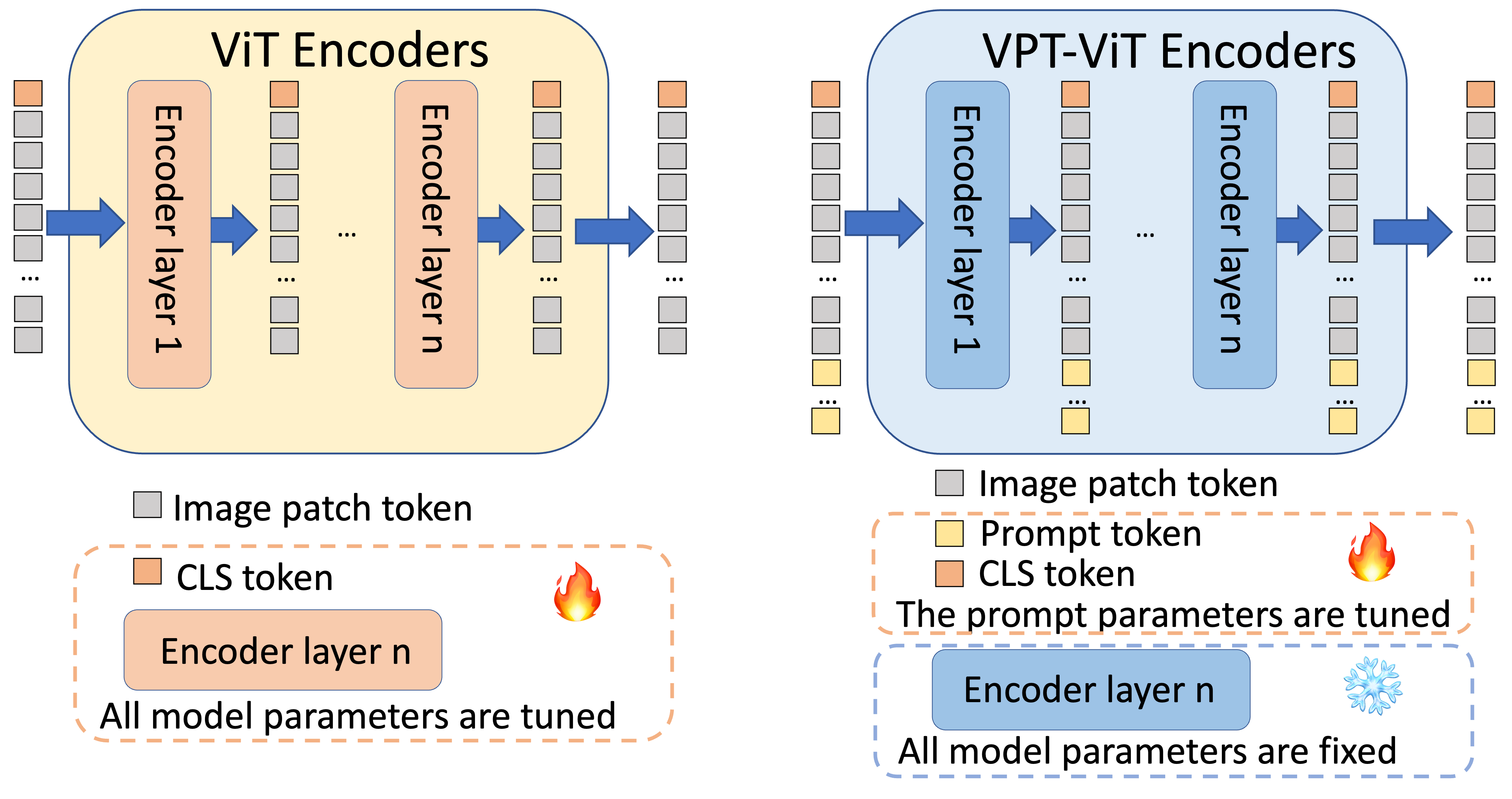}
   \caption{The prompting design of the encoder model. The additional prompt tokens employed explicitly seam the pathological focuses with the backbone’s general vision knowledge. Accordingly, only the prompt tokens are tuned while the other model parameters are frozen.}
   \label{fig:fig_VPT}
\end{figure}

Considering the general vision knowledge, we explicitly use prompt tokens to seam the pathological focuses into the pre-training process. Following the VPT-Deep design in \cite{vpt}, we build VPT-ViT encoders with additional prompt tokens at each encoder layer $n$, as shown in Fig. \ref{fig:fig_VPT}. In this prompt training, only the prompt tokens $P_n$ of VPT are updated to learn the pathological domain features. The remaining ViT parameters carrying the general vision knowledge are frozen. Accordingly, the encoder learning is then formulated as follows.

\begin{equation}
\resizebox{\linewidth}{!}{
    $[T_{b}^{cls,n}, T_{b}^{Pos,n}, T_{b}^{Rel,n}] = L_n\left[T_{b}^{cls,n-1},T_{b}^{Pos,n-1},T_{b}^{Rel,n-1}, P_n \right]$}
\label{eq:encoder_backbone_3}
\end{equation}

The encoder model patches relationships through puzzle understanding and the prompt tokens explicitly learn the domain-bridging knowledge aligning general vision.

\subsubsection{Positional hint}

As shown with purple lines in Fig. \ref{fig:fig_PuzzleTuning_method}, the positional hint is a special design that enables the shuffled puzzle bags to be adequately restored. Specifically, after puzzle understanding, the original position tokens $T_{b}^{Pos,0}$ replace the learned position tokens $T_{b}^{Pos,N_{Enc}}$. They serve as a hint of the puzzle starting point for the subsequent puzzle restoring.

\begin{equation}
  \resizebox{\linewidth}{!}{
    $[T_{b}^{cls,N_{Enc}}, T_{b}^{Pos,N_{Enc}}, T_{b}^{Rel,N_{Enc}}] \rightarrow \left[T_{b}^{cls,N_{Enc}}, T_{b}^{Pos,0}, T_{b}^{Rel,N_{Enc}}\right]$}
  \label{eq:positional_hint_1}
\end{equation}

Combined with the learned relation tokens $T_{b}^{Rel,N_{Enc}}$ carrying the abstract information, each latent puzzle image $\hat{I}_{b}^{latent}$ is obtained by restoring token locations. As the fix-position tokens are randomly assigned, the positional hint is also flexible and only links the position tokens $T_{b}^{Pos,N_{Enc}}$. Following the same sequence of original input images $I_{b}$, the awareness of batch order for the decoder enables SSL loss supervision.

\begin{equation}
  [T_{b}^{Pos,0}, T_{b}^{Rel,N_{Enc}}] \rightarrow \left\{ \hat{I}_{b}^{latent} \in \mathbb{R}^{3,h,w} \right\}
  \label{eq:positional_hint_2}
\end{equation}

\subsection{Puzzle Restoring Process}

The decoder first embeds the latent puzzle image $\hat{I}_{b}^{latent}$ into $k_{Dec}$ tokens as $T_{b}^{Dec}$, following the sequence of latent puzzle images $\{\hat{I}_{b}^{latent} | b=1,\cdots,B\}$ with its corresponding hint $T_{b}^{Pos,0}$ and learned latent tokens $T_{b}^{Rel,N_{Enc}}$. 

\begin{equation}
  T_{b}^{Dec} = \left\{ t_b^k \in \mathbb{R}^{D_{Dec}} \right\}
  \label{eq:puzzle_restoring_1}
\end{equation}

Each token $t_b^k$ is of the $D_{Dec}$ dimension, where $k=1, \cdots, k_{Dec}$, $b=1, \cdots, B$. Due to the decoder design, the feature dimension $D_{Dec}$ can be different from the encoder dimension $D_{Enc}$.
Next, the image $\hat{I}_{b}^{restore}$ is reconstructed with a size of $3,h,w$ pixels.

\begin{equation}
  \hat{I}_{b}^{restore} = Decoder(T_{b}^{Dec})
  \label{eq:puzzle_restoring_2}
\end{equation}

Targeting to reconstruct the permuted relation patches $I_{b}^{Rel}$, the $\hat{I}_{b}^{restore}$ is cropped into a set of $m$ patches (size of $(3, P, P)$ pixels). 
Accordingly, $m$ position tokens $I_{b}^{Pos,restore}$ and $m-r$ relation tokens $I_{b}^{Rel,restore}$ are identified, following the corresponding assignment $x \in \{1, \cdots, m\}$ on the location index of the patches in $I_{b}$. The reconstruction is therefore denoted as $\hat{I}_{b}^{restore} = \{I_{b}^{Pos,restore} \cup I_{b}^{Rel,restore}\}$, where

\begin{equation}
\begin{split}
  I_{b}^{Rel,restore} = \left\{ I_{x,b}^{Rel,restore} \big| x = 1, \dots, m-r \right\}
  \label{eq:puzzle_restoring_3}
\end{split}
\end{equation}

The mean squared error (MSE) loss is applied for this SSL process, which supervises the decoder with the target patches $I_{x,b}^{Rel}$ at location index $x$ and batch index $b$.

\begin{equation}
  loss=MSE(I_{x,b}^{Rel,restore}, I_{x,b}^{Rel})
  \label{eq:puzzle_restoring_4}
\end{equation}


\section{Experiment}
\label{sec:experiment}

In this section, we mainly explore and evaluate the effectiveness of PuzzleTuning.

\subsection{Experiment settings}

\subsubsection{Pre-training dataset}
\begin{figure}[t]
  \centering
   \includegraphics[width=1.0\linewidth]{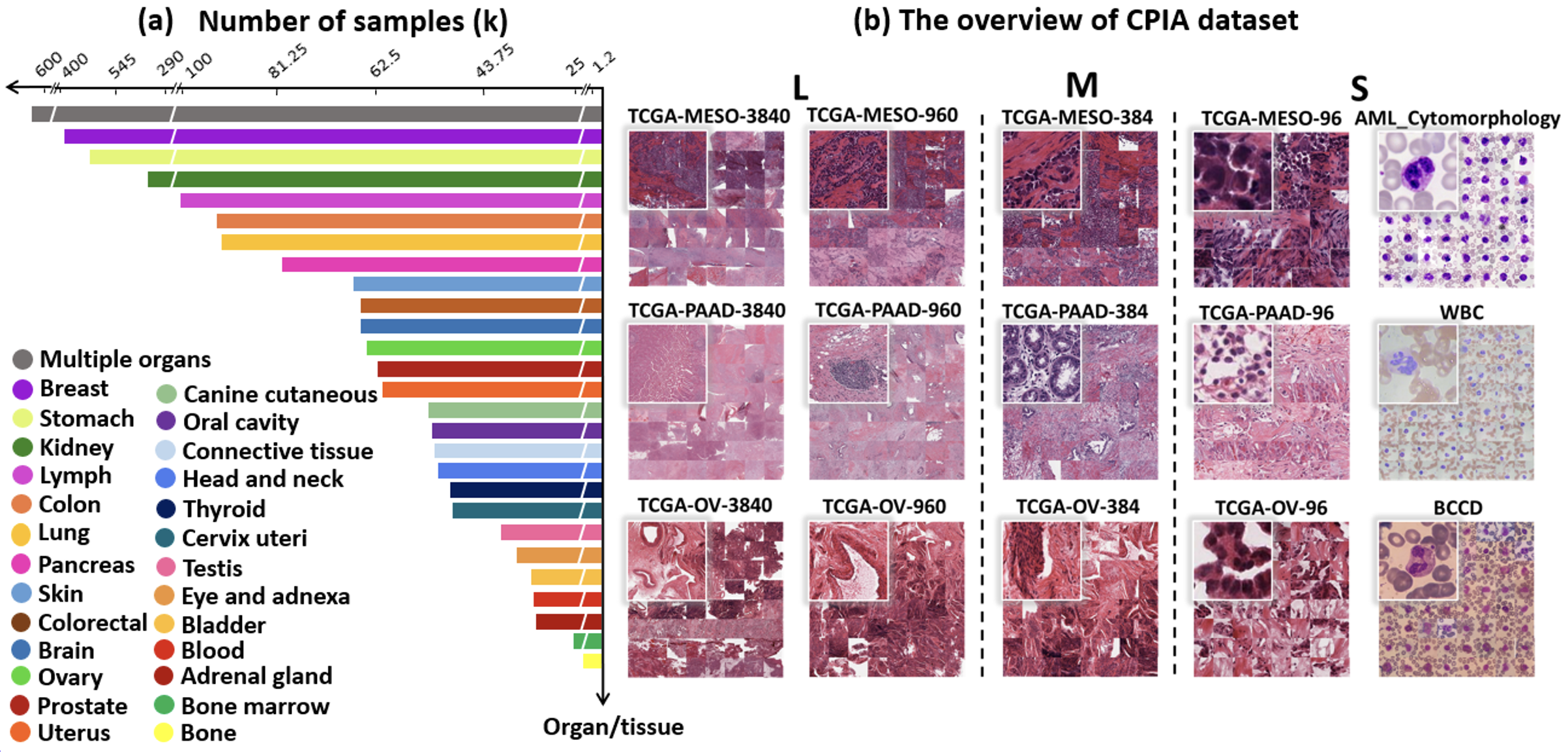}
   \caption{The data distribution and overview of the CPIA-Mini. (a) The organ/tissue categories and sample numbers of the CPIA-Mini dataset. Notably, the multiple organs category indicates that certain sub-datasets contain samples of more than one organ/tissue. (b) The thumbnail of the CPIA dataset. The CPIA-Mini dataset includes three scales reflecting the diagnosis habits of pathologists.}
   \label{fig:fig_dataset}
\end{figure}

Experiments in PuzzleTuning include two steps: pre-training and downstream finetuning. In the first self-supervised pre-training stage, a wide-covering Comprehensive Pathological Image Analysis Dataset (CPIA) \cite{ying2023cpia} is used. Specifically, we utilized the CPIA-Mini dataset, which is a manually selected subset of the CPIA dataset amalgamating a variety of 100 diseases. It consists of 3,383,970 un-labelled images from 48 organs and tissues. Designed for broad coverage, it includes three manually categorized image magnifications (large, medium, and small notated as L, M, and S) by expert pathologists to address the multi-scale nature of pathological image analysis. The data distribution and details of these sub-scales are presented in Fig. \ref{fig:fig_dataset}.

\subsubsection{Comparison SOTA pre-training methods}

To illustrate the effectiveness of PuzzleTuning, we benchmark several SSL SOTAs via equitable pre-training and finetuning. The CL-based SSL methods such as Virchow \cite{virchow} (DINO \cite{emerge_ssl_vit}) and TransPath \cite{transpath} (BYOL \cite{byol}) in the pathological domain, and MoCo-V3 \cite{moco} and SimCLR \cite{simclr} in the general computer vision (CV) field are trained. Then, we evaluated recent MIMs such as MAE \cite{mae}, CAE \cite{CAE}, MaskFeat \cite{MaskFeat} from the general CV field and their specialized counterparts for pathology, namely GC-MAE \cite{gcmae}. Combining with CL objectives, the hybrid methods such as SimMIM \cite{simmim} and BeyondMask \cite{BeyondMask}, and the similar works with the puzzle intention such as Jigsaw \cite{jigsaw}, DropPos \cite{wang2023droppos} are also compared.

\subsubsection{Implementation details}
In pre-training, ViT-base is initialized with ImageNet, and then the different pre-training methods are compared. For each pre-training model, their training hyper-parameters are set following their official release and then further optimized on CPIA-mini. In PuzzleTuning pre-training with VPT, there are 20 prompt tokens attached in each of ViT encoder layer. After 20 epochs of warm-up, 180 epochs are trained with a learning rate of 0.0001 and cosine weight decay to 0.05 times. The batch size is set to 256 images per GPU, and the puzzle group size is 16. The patch-scheduler controls puzzle patch sizes, altering scales looping from 16, 32, 48, 64, 96, and 112. The fix-position ratio scheduler linearly reduces the fixed-patch ratio from 90\% to 20\% to gradually increase the puzzle shuffling complexity. 

After pre-training, on each downstream task, the pre-trained models are finetuned with multiple optimized hyper-parameters. For each output model, the average of the top-5 best-performed experiments are reported. There are 4 A100 SMX4 GPUs used in pre-training and 4 RTX3090 GPUs used in downstream finetuning. More details of experiments are presented in the online release. Additionally, we present a Colab demo online for illustration at: https://github.com/sagizty/PuzzleTuning.

\subsection{Effectiveness on ROI Classification}

\subsubsection{ROI datasets for finetuning}
In the finetuning stage for ROI classification, four diverse datasets are utilized to evaluate the classification performance across multiple pathological scales from cellular to tissue level. These datasets are split in a manner reflective of their content: CAM16 \cite{camelyon}, pRCC \cite{ivit}, and ROSE \cite{zhang2022shuffle} are divided into training, validation, and test sets with a 7:1:2 ratio, while the WBC \cite{white_blood_dataset} dataset, considering its original separation, follows a 10:2:5 ratio. To ensure data integrity and prevent leakage, the ROSE and the test sets for other datasets are independent of CPIA-Mini. In addition to Table \ref{tab:table_dataset}, more details of these datasets are provided as follows:

\begin{table}
    \begin{center}
    \caption{The overview of downstream tasks and datasets. After pre-training, the models are fine-tuned on each task.}
    \label{tab:table_dataset}
    \resizebox{0.47\textwidth}{!}{%
    \begin{tabular}{c|c|c|c|c}
        \toprule
        Dataset & Task & Sample Number & Organ/Tissue & Feature Scale \\
        \midrule
        CAM16 & binary classification & 1081 ROIs & Lymph Nodes & Tissue \\
        pRCC & binary classification & 1417 ROIs & Kidney & Glandular \\
        ROSE & binary classification & 5088 ROIs & Pancreas & Cellular \\
        WBC & 5-class classification & 14514 ROIs & Blood & Subcellular \\
        PanNuke & 5-class segmentation & 7904 ROIs & Multiple & Cellular \\
        lUAD \& LUSC & 3-class WSI classification & 2063 WSIs & Lung & Tissue \\
        \bottomrule
    \end{tabular}
    }
    \end{center}
  
\end{table}

Camelyon16 (CAM16): From the Cancer Metastases in Lymph Nodes challenge, this dataset consists of 540 tumors and 541 normal histopathology images, each cropped to 8000x8000 dimensions.

Papillary Renal Cell Carcinoma Dataset (pRCC): It includes 870 type I and 547 type II histopathology images, with an average size of 2000x2000. Type I images present small cells with clear cytoplasm, while Type II display cells with voluminous cytoplasm and high-grade nuclei.

ROSE Dataset: A private collection from Peking Union Medical College Hospital, this dataset contains cytopathology images from pancreatic liquid samples, including 1,773 pancreatic cancer and 3,315 normal images.

Raabin-WBC Dataset (WBC): Comprising cytopathology images of five blood cell types, it includes 301 basophil, 1,066 eosinophil, 3,461 lymphocyte, 795 monocyte, and 8,891 neutrophil images.

\subsubsection{Classification performance}

\begin{table*}[htbp]
    \begin{center}
    \caption{The average top-5 accuracy and F1 (F1-score) comparison with SOTA pre-training methods on 4 downstream classification datasets including CAM16, pRCC, ROSE, and WBC. P-values of the paired-t test are given for comparison on the F1-score, validating pre-training performance.}
    \label{tab:table_results_main}
    \resizebox{\textwidth}{!}{%
    \begin{tabular}{lcccccccccccc}
    \toprule
    \textbf{Method} & \multicolumn{3}{c}{\textbf{CAM16}} & \multicolumn{3}{c}{\textbf{pRCC}} & \multicolumn{3}{c}{\textbf{ROSE}} & \multicolumn{3}{c}{\textbf{WBC}} \\ \cline{2-13}
                    & \textbf{Acc(\%)} & \textbf{F1(\%)} & \textbf{P-value} & \textbf{Acc(\%)} & \textbf{F1(\%)} & \textbf{P-value} & \textbf{Acc(\%)} & \textbf{F1(\%)} & \textbf{P-value} & \textbf{Acc(\%)} & \textbf{F1(\%)} & \textbf{P-value} \\ \midrule
    SimCLR & 92.69 & 92.68 & 4.28E-06 & 93.43 & 92.89 & 3.43E-02 & 91.85 & 90.93 & 4.43E-04 & 98.11 & 96.82 & 1.37E-03 \\
    MOCO & 86.30 & 86.21 & 6.72E-03 & 75.55 & 72.43 & 1.95E-06 & 72.28 & 63.84 & 3.02E-09 & 91.37 & 86.40 & 3.98E-06 \\
    TransPath (BYOL) & 91.48 & 91.47 & 1.12E-02 & 75.48 & 72.73 & 1.01E-05 & 92.58 & 91.75 & 1.25E-02 & 95.55 & 92.55 & 1.35E-05 \\
    Virchow (DINO) & 82.50 & 82.29 & 7.25E-05 & 74.20 & 71.28 & 2.21E-06 & 79.76 & 76.02 & 3.89E-06 & 90.79 & 85.45 & 1.81E-05 \\
    MAE & 93.33 & 93.33 & 1.72E-04 & 93.85 & 93.48 & 1.37E-01 & 91.22 & 90.19 & 9.58E-06 & 97.88 & 96.21 & 3.98E-04 \\
    CAE & 81.39 & 81.17 & 7.78E-06 & 72.16 & 69.16 & 1.94E-08 & 72.20 & 63.47 & 3.59E-08 & 86.94 & 76.61 & 1.52E-02 \\
    GCMAE & 93.52 & 93.52 & 2.36E-14 & 90.95 & 90.30 & 1.04E-04 & 90.02 & 88.85 & 4.48E-06 & 97.08 & 95.16 & 3.81E-06 \\
    MaskFeat & 94.07 & 94.07 & 4.45E-05 & 92.93 & 92.39 & 1.01E-04 & 90.73 & 89.73 & 1.50E-04 & 97.43 & 95.73 & 5.55E-05 \\
    DropPos & 91.11 & 91.10 & 1.10E-05 & 79.58 & 77.68 & 3.64E-08 & 83.74 & 81.93 & 6.22E-08 & 96.04 & 93.48 & 2.67E-06 \\
    JIGSAW & 88.06 & 88.02 & 2.19E-05 & 73.92 & 70.78 & 2.76E-06 & 73.84 & 65.93 & 1.28E-06 & 92.05 & 87.74 & 8.77E-06 \\
    SIMMIM & 93.52 & 93.52 & 3.56E-04 & 88.62 & 87.62 & 1.58E-06 & 90.53 & 89.39 & 3.43E-06 & 97.03 & 95.03 & 2.14E-05 \\
    BeyondMask & 92.41 & 92.40 & 4.95E-05 & 86.22 & 84.96 & 1.02E-06 & 87.46 & 85.97 & 2.33E-06 & 96.34 & 94.07 & 1.71E-07 \\
    \textbf{PuzzleTuning} & \textbf{95.83} & \textbf{95.83} & \textbf{-} & \textbf{93.92} & \textbf{93.54} & \textbf{-} & \textbf{93.27} & \textbf{92.52} & \textbf{-} & \textbf{98.49} & \textbf{97.36} & \textbf{-} \\
    \bottomrule
    \end{tabular}
    }
    \end{center}

\end{table*}
For a fair comparison, we pre-train and finetune the SOTA methods in the same way as PuzzleTuning, where the ImageNet pre-trained ViT-base is trained with the CPIA-mini to adapt to pathological images. All their output models undergo finetuning with various hyper-parameters across four datasets, where the averaged top-5 results are reported in Table \ref{tab:table_results_main}. Results show that PuzzleTuning consistently outperforms the comparison methods, achieving accuracy improvements of +2.31\% to 14.44\% on CAM16, +0.07\% to 21.76\% on pRCC, +0.69\% to 21.07\% on ROSE, and +0.38\% to 11.55\% on WBC dataset. Generally, recent MIMs exhibit superior results to the CL methods. Among them, PuzzleTuning, through the explicit SSL pre-training task of multiple puzzle restoring, yields the highest performance. Achieving significant numerical improvements across all tasks, the explicit design bridges the general vision knowledge with pathological knowledge effectively, especially considering only prompt tokens are updated.

\subsection{Effectiveness on Cell Segmentation}

In previous sections, we have evaluated the PuzzleTuning pre-trained models on high-level classification tasks. In this section, we further explore downstream low-level tasks on weakly supervised semantic segmentation (WSSS). Specifically, we adopted CellViT \cite{CellViT}, a recent WSSS framework designed for cell segmentation, with ViT as its backbone. A U-Net-like encoder-decoder structure is used with skip connections concatenating each pair of ViT encoder and CNN decoder layers. We have modified the input image size of CellViT from the original 256*256 to 224*224 to match our pre-training ViT. Additionally, we designed CellVPT, which changes the backbone from ViT to VPT by utilizing additional prompt tokens. In both ViT and VPT designs, we evaluate the effectiveness of PuzzleTuning pre-trained models against others. 

Following CellViT \cite{CellViT}, the PanNuke dataset \cite{gamper2020pannuke} is applied, which contains 7,904 images in size of 256*256. Specifically, 189,744 annotated nuclei from 19 different tissue types and 5 distinct cell categories are included. With pre-trained weights, the CellViT and CellVPT models are trained for 160 epochs, with a learning rate of 0.00005 for CellViT and 0.0002 for CellVPT. Each comparison experiment is trained under the same or higher-performed settings. In the ViT experiments, the pre-trained ViT are loaded as the backbone while their VPT loads additional empty prompt tokens. In the PuzzleTuning VPT experiments, baseline ViT is ImageNet-trained ViT, and PuzzleTuning pre-trained VPT prompt tokens are loaded. Prompting strategies are explored in the downstream WSSS, where only the prompt tokens are updated in the '+pt' process while '+ft' has all parameters trained.

\begin{table}[htbp]
    \begin{center}
    \caption{The results of pre-trained weights on the CellViT and CellVPT for WSSS nuclei segmentation, where ViT/VPT denote different backbone models and pt/ft denote updating the prompt tokens or updating all the parameters in downstream training.}
    \label{tab:table_WSSS_comparison}
  \resizebox{0.47\textwidth}{!}{%
    \begin{tabular}{lcccccccc}
    \toprule
    \textbf{Initialization} & \multicolumn{2}{c}{\textbf{ViT+ft}} & \multicolumn{2}{c}{\textbf{ViT+pt}} & \multicolumn{2}{c}{\textbf{VPT+ft}} & \multicolumn{2}{c}{\textbf{VPT+pt}} \\ 
    \cline{2-9} & DICE & Jacard & DICE & Jacard & DICE & Jacard & DICE & Jacard \\
    \midrule
    SimCLR         & 79.37 & 70.97 & 78.99 & 70.88 & 78.80 & 70.29 & 78.99 & 70.88 \\
    MoCo           & 76.60 & 66.97 & 78.13 & 69.09 & 75.81 & 65.72 & 78.11 & 69.08 \\
    BYOL           & 76.33 & 65.87 & 60.42 & 48.13 & 62.11 & 49.84 & 63.21 & 50.61 \\
    DINO           & 77.01 & 67.51 & 77.32 & 67.79 & 75.73 & 65.54 & 77.32 & 67.79 \\
    MAE            & 79.29 & 71.05 & 79.91 & 72.04 & 79.60 & 71.55 & 79.84 & 71.98 \\
    CAE            & 74.97 & 64.58 & 70.37 & 57.98 & 70.60 & 58.39 & 70.35 & 57.96 \\
    GCMAE          & 79.70 & 71.72 & 78.16 & 69.18 & 79.41 & 71.37 & 78.16 & 69.18 \\
    MaskFeat       & 80.01 & 72.13 & 79.81 & 71.85 & 79.35 & 71.19 & 79.83 & 71.92 \\
    DropPos        & 78.94 & 70.45 & 74.94 & 64.26 & 77.53 & 68.36 & 75.02 & 64.40 \\
    Jigsaw         & 77.30 & 67.63 & 75.75 & 65.42 & 75.02 & 64.20 & 75.81 & 65.47 \\
    SimMIM         & 79.06 & 70.48 & 79.49 & 71.00 & 77.92 & 68.77 & 79.49 & 70.95 \\
    BeyondMask     & 79.56 & 71.46 & 76.25 & 66.10 & 74.34 & 63.52 & 76.29 & 66.07 \\
    \textbf{PuzzleTuning}   & \textbf{79.80} & \textbf{71.98} & \textbf{80.05} & \textbf{72.43} & \textbf{79.80} & \textbf{71.96} & \textbf{79.97} & \textbf{72.18} \\
    \bottomrule
    \end{tabular}
    }
    \end{center}
    
\end{table}

Explicitly enhanced by the pre-training with explicit tasks focusing on appearance consistency, spatial consistency, and restoration understanding, PuzzleTuning further improved the low-level vision tasks for the ViT backbone. As shown in Table \ref{tab:table_WSSS_comparison}, PuzzleTuning pre-trained ViT and VPT achieve the DICE score of 79.80, 80.05, 79.80, 79.97, through the finetuning and prompting process of WSSS. The improved results of PuzzleTuning regarding other SSL pre-training methods support the effectiveness of our explicit task design in low-level downstream tasks.

\subsection{Effectiveness on Whole Slide Images Classification}

To further evaluate the effectiveness of PuzzleTuning pre-training in pathological image analysis, we explored a lung cancer sub-typing task using whole slide images (WSIs) for slide-level classification. Similar to the region-of-interest (ROI) images in other sections, WSIs are commonly used diagnostic tools \cite{lu2021ai}. They are of gigapixel scale and cannot be directly processed by GPUs. Most studies crop each WSI into thousands of patches. However, the scarcity of labels results in only having sample-level labels instead of patch-level ones. Given these characteristics, SSL pre-training markedly enhances the performance of WSI models. 

Specifically, most WSI studies involve a two-stage process: a feature extraction stage, where the cropped patches are embedded into features, and a feature modeling stage, where a model is employed to analyze these extracted features. Accordingly, we employed the pre-trained ViT into two distinct roles with two widely applied WSI analysis frameworks: CLAM \cite{clam} and Graph-Transformer (GTP) \cite{gtp}. Additionally, we encompass the prompt tokens in several explorations, where the VPT is applied with the pre-trained ViT backbone and prompt tokens. 

In this study,  789 lung adenocarcinoma (LUAD),  707 lung squamous cell carcinoma (LUSC), and 567 non-cancerous (normal) tissue WSI samples from TCGA-LUAD and TCGA-LUSC \cite{weinstein2013cancer} datasets are used. The WSIs are separated into training, validation, and test sets based on patient ID with a ratio of 7:1:2. All models are trained 50 epochs with the same learning rate of 0.00005 or hyper-parameters optimized for higher performance.

\begin{table}[htbp]
    \begin{center}
    \caption{The accuracy of PuzzleTuning pre-trained model in WSI slide-level classification. Serving in the feature extraction stage of CLAM \cite{clam}, pre-trained model embeds patches into features; Serving in the feature modeling stage of GTP \cite{gtp}, pre-trained models predict slide-level category with the embedded WSI features. ViT: ViT-base model, VPT: prompting with additional prompt tokens of ViT. In the downstream tasks: ft: finetuning all parameters, pt: prompting only the prompt tokens. Two PuzzleTuning curriculum designs are explored with p16-rd: patch size is fixed at 16 and fix-position ratio decays from 90\% to 20\%, p16-r25: patch size and fix-position ratio are fixed at 16, 25\%.}
    \label{tab:table_wsi}
  \resizebox{0.47\textwidth}{!}{%
    \begin{tabular}{lcccccc}
    \toprule
    \textbf{Initialization} & \multicolumn{2}{c}{\textbf{Feature Extraction}} & \multicolumn{4}{c}{\textbf{Feature Modeling}} \\ 
    \cline{2-7} & ViT & VPT & ViT+ft & ViT+pt & VPT+ft & VPT+pt \\
    \midrule
    Random           & 53.99 & --    & 73.93 & 76.69 & 72.70 & 76.69 \\
    ImageNet         & \textbf{82.82} & --    & 76.07 & 77.91 & 76.69 & 77.91 \\
    SimCLR           & 65.34 & --    & 76.07 & 74.54 & 77.30 & 74.54 \\
    MoCo             & 53.68 & --    & 74.85 & 77.31 & 76.07 & 77.31 \\
    BYOL             & 55.21 & --    & 73.93 & 73.62 & 75.77 & 73.62 \\
    DINO             & 54.60 & --    & 76.07 & 76.69 & 76.07 & 76.69 \\
    MAE              & 57.06 & --    & 75.77 & 75.46 & 77.91 & 75.46 \\
    CAE              & 53.68 & --    & 73.62 & 74.85 & 75.77 & 74.85 \\
    GCMAE            & 61.35 & --    & 74.23 & 75.46 & 76.69 & 75.46 \\
    MaskFeat         & 64.72 & --    & 73.01 & 74.54 & 76.07 & 74.54 \\
    DropPos          & 58.90 & --    & 73.62 & 74.23 & 75.46 & 74.23 \\
    Jigsaw           & 52.15 & --    & 75.46 & 77.30 & 75.46 & 77.30 \\
    SimMIM           & 55.83 & --    & 74.85 & 77.91 & 76.69 & 77.91 \\
    BeyondMask       & 64.42 & --    & 75.15 & 77.30 & 76.99 & 77.30 \\
    PuzzleTuning-p16r25 & 64.42 & 77.91 & 76.07 & 76.69 & 74.23 & 78.83 \\
    PuzzleTuning-p16rd  & \textbf{74.85} & 80.67 & \textbf{76.69} & 77.61 & 76.69 & 75.15 \\
    \textbf{PuzzleTuning}     & 66.56 & \textbf{87.42} & 76.38 & \textbf{78.22} & \textbf{79.14} & \textbf{79.45} \\
    \bottomrule
    \end{tabular}
    }
    \end{center}
    
\end{table}

\subsubsection{In WSI feature extraction}

Applied in the feature extraction stage, PuzzleTuning-trained ViTs are employed in CLAM \cite{clam} to embed cropped images into feature vectors of 768 dimensions. Subsequently, the CLAM framework is trained with the embedded feature vectors for slide-level classification. As depicted in the first two columns of Table \ref{tab:table_wsi}, the performance of the comparisons varies significantly due to the fixed feature extractor, which is not updated during downstream training. Among them, the PuzzleTuning-trained VPT achieves the highest performance with an accuracy of 87.42\%. Furthermore, compared to most other pre-trained models, PuzzleTuning pre-trained ViT models demonstrate superior performance.

However, the highest performance with ViT backbone alone is observed with ImageNet's natural knowledge initialization (accuracy of 82.82\%). As the pre-trained ViTs underperform compared to the ImageNet baseline without pathological image pre-training, the drop in embedding performance (ranging from -16.26 to -30.67\%) further underscores the importance of general vision knowledge. Nevertheless, through prompt tuning in PuzzleTuning, additional domain-bridging knowledge can be effectively encoded, resulting in a performance increase (82.82 to 87.42\%). Moreover, explicit bridging yields significant performance improvements (87.42 vs. 66.56\%) over implicit finetuning of all parameters. Its efficacy at the WSI feature extraction stage supports our objective of adapting natural to the pathological images by learning the bridging prompts.

\subsubsection{In WSI feature modeling}

Applied in the feature modeling stage, GTP \cite{gtp} is built with the ViT/VPT  backbone for slide-level classification. In their training process, the GTP models are initialized with ViT weights from the pre-training stage, and if the VPT is used, the empty additional prompt tokens are attached. For the PuzzleTuning experiments, the ViT+pt loads pre-trained ViT and empty prompt tokens, while the VPT+pt loads baseline ViT and PuzzleTuning pre-trained prompt tokens. In the '+ft' experiments, the ViT and VPT have all parameters trained, while only the prompt tokens are updated in the '+pt' process. 

In the right four columns of Table \ref{tab:table_wsi}, the PuzzleTuning pre-trained models achieve the slide-level classification accuracy of 76.38\%, 78.22\%, 79.14\%, and 79.45\%. Consistent with other comparison experiments against the SOTAs, applying PuzzleTuning-trained models achieve the highest performance in the WSI feature modeling stage. In the slide-level downstream task, the relationships of distinct features in the WSI are effectively modeled. This further proves the effectiveness of multiple puzzle restoration tasks for building relationships based on appearance and spatial consistency. Additionally, the findings in WSI applications further enhance its ability as a general approach in pathological image analysis.

\subsection{Reconstruction Performance}

\begin{figure}[t]
  \centering
   \includegraphics[width=1.0\linewidth]{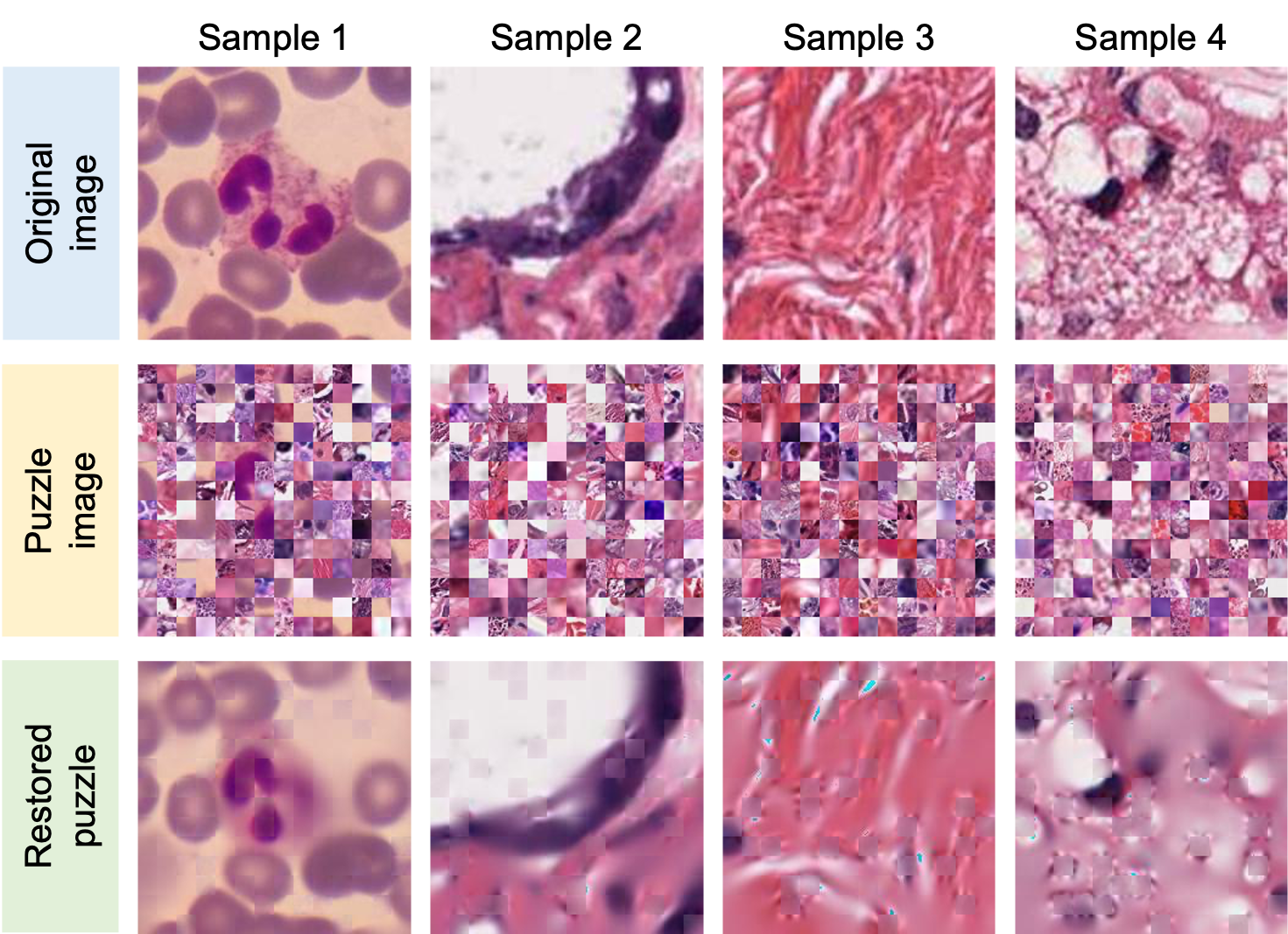}
   \caption{Samples and their permuted and restored puzzles by our proposed PuzzleTuning, illustrating the SSL pre-training task of multiple puzzle restoring.}
   \label{fig:fig_reconstruction_performance}
\end{figure}

\begin{figure}[t]
  \centering
   \includegraphics[width=1.0\linewidth]{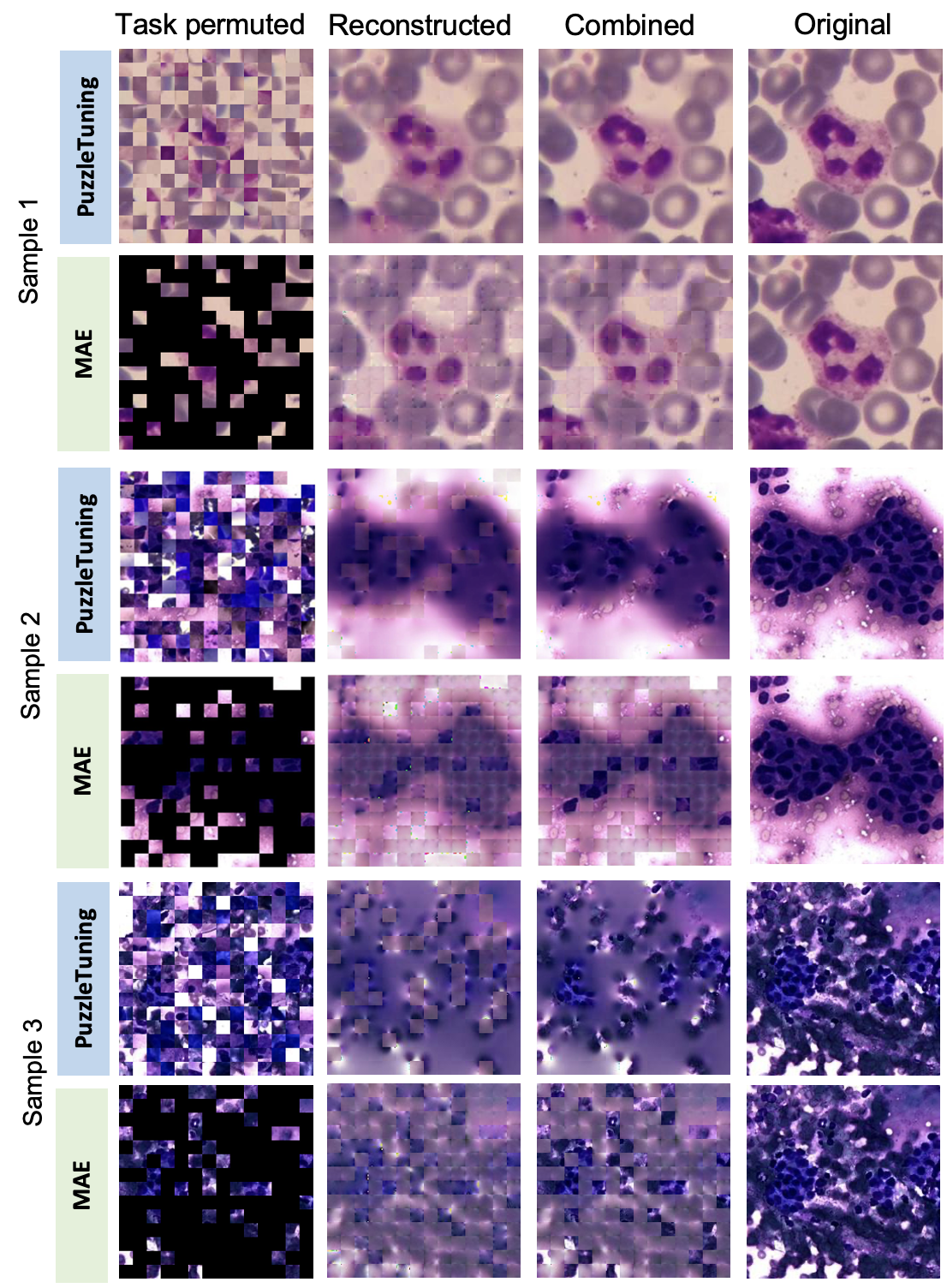}
   \caption{Illustrations of the permuted (masked by MAE or puzzled by PuzzleTuning) samples and reconstructed images. The combined image is generated with the visible patches (the unmasked patches or hint patches) replaced by the original patches, illustrating the reconstruction consistency between patches.}
   \label{fig:fig_reconstruction_comparison}
\end{figure}

We illustrate the original images, puzzles, and reconstructions in PuzzleTuning training in Fig. \ref{fig:fig_reconstruction_performance}. Regarding the SSL task to understand pathological images, both MAE and PuzzleTuning are trained through reconstruction. However, the task focuses differ in their learning and restoring approaches, where PuzzleTuning employs multiple puzzles restoration to explicitly learn the grouping, junction, and semantic alignment relationship. In contrast, MAE mainly focuses on interpreting the junction relationship of remaining patches with MIM. 
With the explicit task design on appearance consistency, spatial consistency, and restoration understanding of Fig. \ref{fig:fig_concept}, we visually compare PuzzleTuning with MAE. In Fig. \ref{fig:fig_reconstruction_comparison}, guided by 25\% of visible original patches, a batch of 4 images and their puzzles and masked samples are generated.

\begin{figure}[t]
  \centering
   \includegraphics[width=1.0\linewidth]{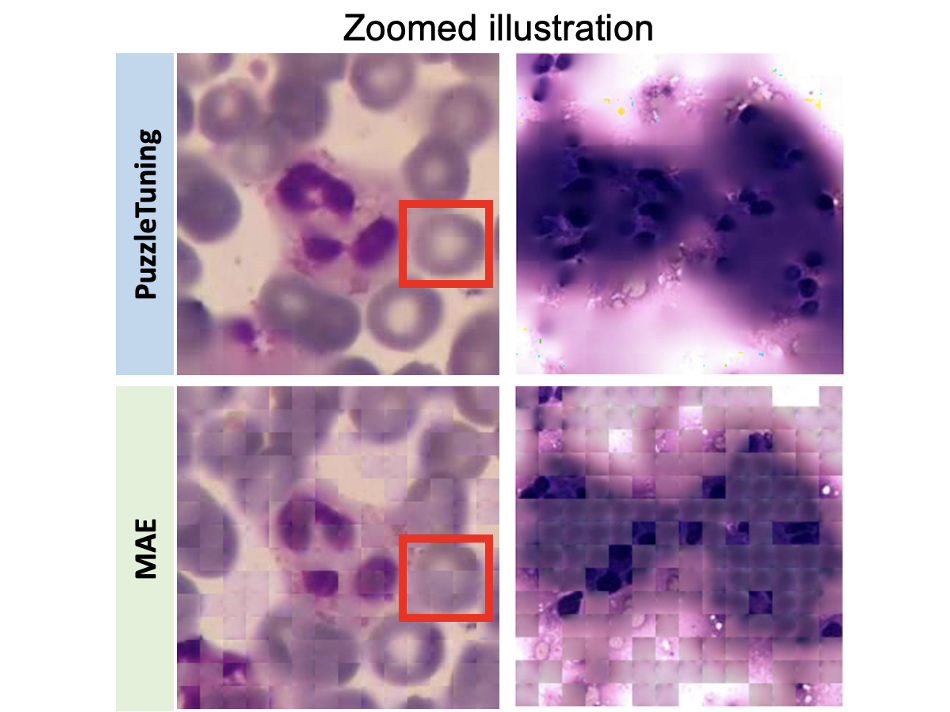}
   \caption{Illustrations of the reconstructed images (by MAE and PuzzleTuning) with the visible patches (the unmasked patches or hint patches) replaced by the original patches. The combined image of PuzzleTuning closely resembles the original image with a low-saturation purple color feature, presenting a smooth and coherent appearance. In contrast, the MAE-generated patches display significant color differences from the original patches.}
   \label{fig:fig_reconstruction_comparison_zoomed}
\end{figure}

Regarding appearance consistency, PuzzleTuning demonstrates a superior capability to reconstruct homogeneous global and local features compared to MAE. Globally, on the first image set of blood cells (sample 1) in Fig. \ref{fig:fig_reconstruction_comparison}, the combined image of PuzzleTuning closely resembles the original image with a low-saturation purple color feature, presenting a smooth and coherent appearance. In contrast, the MAE-generated patches display significant color differences from the original patches. Locally, highlighted in \ref{fig:fig_reconstruction_comparison_zoomed} with red boxes, PuzzleTuning consistently imparts the concave shape in the center of all surrounding red blood cells. In contrast, the MAE-generated images fail to preserve the concave shape of most red blood cells. Similarly, in the second set (sample 2), PuzzleTuning generates cells based on limited visible original patches, restoring cell clusters from the original image. However, MAE-generated images do not effectively restore cell morphology around visible patches, thus failing to achieve the desired consistency between cells.

Regarding spatial consistency presented with junction patches, PuzzleTuning achieves smooth image reconstruction, preserving the advanced texture alignment toward visible original patches. On the contrary, the images generated by MAE exhibit relatively coarse connections between patches. Shown explicitly in the second and the third image sets (sample 2 and 3) in Fig. \ref{fig:fig_reconstruction_comparison}, the edges between different patches in MAE-generated images become notably prominent. Their colors and textures present misalignment to the visible original patches. Additionally, the expected junction textures, such as rounded cell shape, which should be interpreted through the junction patches, are fuzzy in MAE-generated patches in Fig. \ref{fig:fig_reconstruction_comparison}. In contrast, PuzzleTuning reconstructs junction patches more clearly with continuous cell texture and color. Aligned closely with visible patches, PuzzleTuning clearly maintains spatial consistency through smooth transitions along patch edges.
\section{Discussion}
\label{sec:discussion}

PuzzleTuning aims to explicitly seam natural and pathological images with explicit pathological focuses. To obtain the novel objectives, the specific designs in PuzzleTuning are further revealed: (A) the domain-bridging objective: determining which general vision knowledge to be bridged; (B) the SSL task of multiple puzzle restoration: assessing the explicit focus on modeling pathological characteristics; (C) the prompt tokens: evaluating the bridging with additional tokens; (D) the curriculum learning: improving bridging by encoding multiple feature scales and adapting to different complexities gradually.

\subsection{Domain-bridging Objective}

Pathological image analysis has benefited from transfer learning with general vision knowledge. However, the knowledge scope of the natural domain surpasses that of the pathological domain significantly. Therefore, PuzzleTuning seeks to merge the most effective knowledge aspects, and two critical focuses from natural images have caught our interest: low-level semantic knowledge and high-level abstract vision.

Specifically, in Table \ref{tab:table_domain_bridging} and Table \ref{tab:table_WSSS_ablation}, low-level semantic knowledge is obtained through occlusion-invariant learning by MAE \cite{Kong_2023_CVPR}. Conversely, high-level abstract vision knowledge is obtained through traditional supervised classification training (timm). By using their official ViT weights trained on ImageNet, we term these models as 'Semantic' and 'Abstract'. Additionally, we compare them with random initialization, denoted as 'Random'. Then, we employ PuzzleTuning to pre-train those models on pathological images, appending 'PuzzleTuning' before their names. Lastly, after pre-training, we apply the output models by finetuning them on the downstream datasets.

\begin{table}[htbp]
    \begin{center}
    \caption{Average top-5 performance on 4 datasets, with different initialization (Random: randomly set weight of ViT; Semantic: MAE-trained ViT with ImageNet; Abstract: supervised ImageNet-trained ViT), with PuzzleTuning in their names: means applying PuzzleTuning to train the prompt tokens.}
    \label{tab:table_domain_bridging}
  \resizebox{0.47\textwidth}{!}{%
    \begin{tabular}{lcccccccc}
    \toprule
    \textbf{Initialization} & \multicolumn{2}{c}{\textbf{CAM16}} & \multicolumn{2}{c}{\textbf{pRCC}} & \multicolumn{2}{c}{\textbf{ROSE}} & \multicolumn{2}{c}{\textbf{WBC}} \\ 
    \cline{2-9} & Acc(\%) & F1(\%) & Acc(\%) & F1(\%) & Acc(\%) & F1(\%) & Acc(\%) & F1(\%) \\
    \midrule
    Random         & 80.93 & 80.67 & 70.39 & 65.65 & 71.71 & 62.88 & 90.28 & 84.09 \\
    Semantic       & 93.70 & 93.70 & 90.39 & 89.86 & 90.02 & 88.68 & 96.62 & 94.66 \\
    Abstract    & 94.26 & 94.26 & 91.59 & 91.11 & 92.52 & 91.64 & 97.25 & 95.45 \\
    \hline
    PuzzleTuning-Random      & 93.52 & 93.52 & 92.65 & 92.09 & 92.28 & 91.42 & 97.64 & 96.10 \\
    PuzzleTuning-Semantic    & 93.33 & 93.33 & 91.73 & 91.07 & 91.97 & 91.12 & 97.79 & 96.28 \\
    \textbf{PuzzleTuning-Abstract} & \textbf{95.83} & \textbf{95.83} & \textbf{93.92} & \textbf{93.54} & \textbf{93.27} & \textbf{92.51} & \textbf{98.49} & \textbf{97.36} \\
    \bottomrule
    \end{tabular}
    }
    \end{center}
\end{table}

In the high-level vision task of classification, we first discern the type of knowledge initialization. As depicted in Table \ref{tab:table_domain_bridging}, across all four downstream classifications, predictably, random initialization shows the worst performance due to the absence of any prior knowledge. Then, the semantic knowledge group of occlusion-invariant learning via MAE obtains improved performance in downstream classification. Lastly, with further improved performance (+0.56\%, 1.2\%, 2.5\%, 0.63\%, in Acc on the four datasets), abstract initialization emerges as the best-performing candidate.

Then, we explore domain bridging by applying PuzzleTuning on pathological images with these three model initializations. In Table \ref{tab:table_domain_bridging}, generally, all initializations exhibit significant improvements after PuzzleTuning (Accuracy improvement ranging from +7.36\% to +22.26\%, -0.37\% to +1.95\%, and +0.75\% to +2.33\% for the three initializations on the four tasks). The results support the intention of explicitly bridging natural and pathological domains with pathological focuses. Meanwhile, the abstract initialization, with its most prominent performance, turns out to be the most suitable general vision knowledge from the natural images. However, while supervised classification training with ImageNet aligns more closely with the downstream classification task, the efficacy of bridging semantic domain knowledge warrants further investigation with low-level tasks.

\begin{table}[htbp]
    \begin{center}
    \caption{The nuclei segmentation performance with different pre-knowledge in PuzzleTuning pre-training, where the CellViT and CellVPT are used. The ViT/VPT denotes different backbone models and pt/ft denotes updating the prompt tokens or updating all the parameters in downstream training. The knowledge applied is denoted as: Random: randomly set weight of ViT; Semantic: MAE-trained ViT with ImageNet; Abstract: supervised ImageNet-trained ViT. Two PuzzleTuning curriculum are explored with p16-rd: patch size is fixed at 16 and fix-position ratio decays from 90\% to 20\%, p16-r25: patch size and fix-position ratio are fixed at 16, 25\%.}
    \label{tab:table_WSSS_ablation}
  \resizebox{0.47\textwidth}{!}{%
    \begin{tabular}{lcccccccc}
    \toprule
    \textbf{Initialization} & \multicolumn{2}{c}{\textbf{ViT+ft}} & \multicolumn{2}{c}{\textbf{ViT+pt}} & \multicolumn{2}{c}{\textbf{VPT+ft}} & \multicolumn{2}{c}{\textbf{VPT+pt}} \\ 
    \cline{2-9} & DICE & Jacard & DICE & Jacard & DICE & Jacard & DICE & Jacard \\
    \midrule
    Random & 77.18 & 67.76 & 77.18 & 67.74 & 75.55 & 65.23 & 77.15 & 67.64 \\
    Semantic & 77.95 & 68.97 & 79.92 & 71.86 & 79.56 & 71.45 & 79.92 & 71.86 \\
    Abstract & 79.78 & 71.54 & 79.95 & 72.14 & 79.71 & 71.83 & 79.82 & 71.93 \\
    \hline
    PuzzleTuning-p16r25 & 79.14 & 70.81 & 77.21 & 67.75 & 79.69 & 71.69 & 79.87 & 71.97 \\
    PuzzleTuning-p16rd & 78.74 & 70.33 & 78.73 & 69.88 & 79.87 & 71.90 & 79.87 & 71.91 \\
    \hline
    PuzzleTuning-Abstract & 79.80 & 71.98 & \textbf{80.05} & \textbf{72.43} & 79.80 & 71.96 & \textbf{79.97} & \textbf{72.18} \\
    PuzzleTuning-Semantic & \textbf{80.18} & \textbf{72.64} & 79.96 & 72.09 & \textbf{80.00} & \textbf{72.14} & 79.93 & 71.99 \\
    \bottomrule
    \end{tabular}
    }
    \end{center}
    
\end{table}

In low-level vision tasks at Table \ref{tab:table_WSSS_ablation}, comparing abstract (ImageNet) with semantic (MAE-ImageNet) knowledge initialization, they generally yield similar DICE performance (+1.83\%, +0.03\%, +0.25\% and -0.1\%) while abstract initialization still perform slightly better. Consistently, both are significantly better performance than the random knowledge initialization. Encompassed with PuzzleTuning, bridging the abstract general knowledge yields consistent improvement in downstream tasks. Specifically, when bridging the pathological images with semantic (MAE-ImageNet) initialization, the performances gain further improvement in the low-level WSSS tasks (ViT+ft and VPT+ft).

\subsection{Explicit Domain-bridging with Puzzles}

Compared with the SSL MIM methods of MAE, GCMAE, and SimMIM in Table \ref{tab:table_results_main}, PuzzleTuning achieves significantly higher performance in all downstream tasks. With the proposed multiple puzzle restoring task, the additional abstract focuses of appearance consistency and restoration understanding are effectively introduced. These explicit focuses improve the SSL on pathological samples compared to occlusion invariant learning of MIM \cite{Kong_2023_CVPR}, which only targets spatial consistency rooted in junction relationships. 

To further explore this task, an ablation model of PuzzleTuning is specially designed by using the masking task from MAE. This counterpart has relation tokens removed in the encoder (ViT), and its decoder is designed to predict the missing tokens. In Table \ref{tab:table_puzzles_and_prompts}, this ablation version is denoted as MAE, and the standard multiple puzzle restoring version is denoted as Shuffled Auto-Encoder (SAE). The numerical comparison underscores a general uptrend in the SAE over their MAE counterparts (+1.48\%, +4.09\%, +3.29\%, +1.11\% improvements in Acc from their best performed settings in the foir tasks). Results suggest that, with explicit bridging focuses, the puzzle task is more effective than masking.

\begin{table}[htbp]
    \begin{center}
    \caption{Average top-5 performance on 4 datasets, with different tuning methods (In pre-training: SAE: multiple puzzle restoring task proposed with PuzzleTuning, MAE: MAE method; ViT: ViT-base model, VPT: prompting with additional prompt tokens of ViT. In the downstream tasks: ft: finetuning all parameters, pt: prompting only the prompt tokens).}
    \label{tab:table_puzzles_and_prompts}
  \resizebox{0.47\textwidth}{!}{%
    \begin{tabular}{lcccccccc}
    \toprule
    \textbf{Tuning Methods} & \multicolumn{2}{c}{\textbf{CAM16}} & \multicolumn{2}{c}{\textbf{pRCC}} & \multicolumn{2}{c}{\textbf{ROSE}} & \multicolumn{2}{c}{\textbf{WBC}} \\ 
    \cline{2-9} & Acc(\%) & F1(\%) & Acc(\%) & F1(\%) & Acc(\%) & F1(\%) & Acc(\%) & F1(\%) \\
    \midrule
    \textbf{SAE-VPT+ft}   & \textbf{95.83} & \textbf{95.83} & 93.92 & 93.54 & \textbf{93.27} & \textbf{92.51} & 98.49 & 97.36 \\
    SAE-VPT+pt   & 94.17 & 94.17 & 90.74 & 90.08 & 92.01 & 91.09 & \textbf{98.54} & \textbf{97.38} \\
    SAE-ViT+ft   & 95.46 & 95.46 & \textbf{95.19} & \textbf{94.84} & 91.36 & 90.28 & 98.06 & 96.73 \\
    SAE-ViT+pt   & 92.13 & 92.12 & 82.26 & 80.41 & 87.40 & 85.93 & 97.34 & 95.58 \\
    \hline
    MAE-VPT+ft   & \textbf{94.35} & \textbf{94.35} & \textbf{91.10} & \textbf{90.36} & \textbf{89.98} & \textbf{88.84} & 96.72 & 94.80 \\
    MAE-VPT+pt   & 91.02 & 91.01 & 87.28 & 85.96 & 89.86 & 88.53 & \textbf{97.43} & \textbf{95.67} \\
    MAE-ViT+ft   & 91.94 & 91.94 & 87.00 & 85.86 & 86.38 & 84.43 & 96.56 & 94.71 \\
    MAE-ViT+pt   & 91.02 & 91.02 & 82.33 & 80.89 & 85.77 & 83.72 & 96.67 & 94.42 \\
    \bottomrule
    \end{tabular}
    }
    \end{center}
\end{table}

Reconstructions in Fig. \ref{fig:fig_reconstruction_comparison} are visualized with the same 25\% visible patches (the unmasked patches in MAE or hint tokens in PuzzleTuning). The images regenerated by PuzzleTuning bear a closer resemblance to their original appearances. The textures adjacent to the revealed patches are clearer, reflecting the junction relationships internalized by both methodologies. However, compared with SAE, MAE over-focuses on restoring the surrounding details of visible patches and struggles to effectively restore distant regions. Moreover, puzzle reconstructions display better color consistency regarding the original patch tokens, further validating its proficiency in assimilating the grouping relationship with its task focus of appearance consistency. 

\subsection{Explicit Domain-bridging with Prompt}

Table \ref{tab:table_domain_bridging} and previous sections underscore the efficacy of general vision knowledge, which provides a robust foundation for small dataset modeling. However, the differences between the pathological and natural images highlight misalignment in knowledge scope and sample variety. It calls for effective methods to combine domain-specific understanding and maintain generalized abilities for catering transfer learning to specialized tasks.

The utilization of prompt tokens stands out as an explicit strategy for conveying bridging knowledge while keeping the model backbone—representing general vision knowledge—unaltered. As delineated in Table \ref{tab:table_WSSS_ablation}, Table \ref{tab:table_puzzles_and_prompts} and the right feature modeling of Table \ref{tab:table_wsi}, during pre-training, the VPT only turns the auxiliary prompt tokens, integrating puzzle understandings, whereas ViT finetunes the entire ViT backbone. As for the downstream phase, the 'pt' tag denotes only updating prompt tokens, and 'ft' represents updating the whole structure. Encountering the pre-trained ViT backbone, the downstream 'pt' combines the entire trained structure with empty additional prompt tokens. Conversely, if the pre-trained components are solely prompt tokens (‘VPT’), the downstream model initializes the backbone ViT structure using timm weight and then attaches these pre-trained prompt tokens.

In classification tasks, both ROI (Table \ref{tab:table_puzzles_and_prompts}) and the WSI (feature modeling section in Table \ref{tab:table_wsi}) show that the 'VPT' groups surpass the 'ViT' groups. It indicates that tuning the explicit prompt tokens during pre-training tends to yield better performance. Therefore, this observation strengthens the explicit design of seaming different knowledge, where the additional prompt tokens explicitly learn bridging knowledge.
Furthermore, as evidenced in Table \ref{tab:table_wsi}, explicit domain bridging significantly enhances feature extraction performance (87.42\% compared to 66.56\%), where other fine-tuning-based approaches failed to improve upon the baseline ImageNet initialization.
Lastly, in the WSSS task reported in Table \ref{tab:table_WSSS_ablation}, all PuzzleTuning ablation models achieve similar performance. Comparing the 'VPT' versions with 'ViT' versions, pre-trained prompt tokens generally introduce better performance. This consistent finding across experiments in both high-level abstract vision tasks (ROI and WSI classification) and low-level tasks (WSSS) evidently supports our explicit domain bridging intention. Accordingly, by training these bridging prompt tokens on an extensive pathological dataset such as CPIA, models can achieve significant improvements in downstream tasks.

Focusing on the prompt strategies, in classification tasks, finetuning all parameters in downstream tasks generally achieves better performance ('ft' groups outshine the 'pt' groups). Due to the narrower knowledge scope and limited sample variety in specialized downstream datasets, finetuning the entire backbone allows for more precise calibration of both general and bridging knowledge. It ensures better fitting on small pathological datasets. Based on insights from both pre-training and downstream finetuning stages, the most efficacious setup emerges: adjust prompt tokens during pre-training and subsequently finetune both the backbone and prompt tokens in the downstream phase (VPT+ft). 

Conversely, in downstream WSSS at Table \ref{tab:table_WSSS_ablation}, most experiments indicate that updating only the prompt tokens (+pt) achieves better performance than updating the backbone ViT and prompts (+ft). While in the high-level abstract vision tasks, updating both backbone and prompt tokens yields better performance. These findings may related to the task complexity of WSSS, where updating more backbone parameters may encounter slightly more overfitting.

\subsection{Curriculum-based Domain-bridging}

In pathological image analysis, multi-scale semantic knowledge is crucial. PuzzleTuning addresses this by varying puzzle patch sizes regulated by a patch-scheduler. Specifically, we currently employ a repetitive-looping strategy shown in Fig. \ref{fig:fig_curriculum_learning}, cycling patch sizes from 16 to 112 every three epochs. Additionally, inspired by curriculum learning to improve the convergence, a fix-position ratio scheduler is designed, varying the proportion of un-shuffled patches from 90\% to 20\%. It modulates shuffling complexity and thus controls the task difficulty. With the two scheduler design, this standard curriculum is applied to all PuzzleTuning experiments.

\begin{table}[htbp]
    \begin{center}
    \caption{Average top-5 performance on 4 datasets, with different PuzzleTuning curriculums. (base: patch size loop over the patches, and fix-position ratio decay from 90\% to 20\%; p16-rd: patch size is fixed at 16 and fix-position ratio decays from 90\% to 20\%, p16-r25: patch size and fix-position ratio are fixed at 16, 25\%).}
    \label{tab:table_curriculum}
  \resizebox{0.47\textwidth}{!}{%
    \begin{tabular}{lcccccccc}
    \toprule
    \textbf{Curriculum} & \multicolumn{2}{c}{\textbf{CAM16}} & \multicolumn{2}{c}{\textbf{pRCC}} & \multicolumn{2}{c}{\textbf{ROSE}} & \multicolumn{2}{c}{\textbf{WBC}} \\ 
    \cline{2-9} & Acc(\%) & F1(\%) & Acc(\%) & F1(\%) & Acc(\%) & F1(\%) & Acc(\%) & F1(\%) \\
    \midrule
    \textbf{base}              & \textbf{95.59} & \textbf{95.58} & 93.31 & 92.85 & \textbf{93.14} & \textbf{92.39} & 98.25 & 97.00 \\
    p16-rd & 95.22 & 95.21 & \textbf{93.92} & \textbf{93.48} & 92.64 & 91.83 & 98.38 & 97.27 \\
    p16-r25     & 95.06 & 95.06 & 93.29 & 92.77 & 92.54 & 91.72 & \textbf{98.45} & \textbf{97.27} \\
    \bottomrule
    \end{tabular}
    }
    \end{center}
\end{table}

Furthermore, additional curriculum ablations are explored in Table \ref{tab:table_curriculum}, Table \ref{tab:table_wsi} and Table \ref{tab:table_WSSS_ablation}. The standard strategy are denoted as 'base' and  'p16-rd' and 'p16-r25' are designed with fixed patch sizes of 16 but differing fix-position ratios. 
In ROI classification (Table \ref{tab:table_curriculum}), all three curriculum ablations surpass other SOTA methods in Table \ref{tab:table_results_main} significantly. Among them, the 'base' performs best in CAM16 and ROSE datasets, the 'p16-rd' excels in pRCC (+0.61\% in accuracy), and the 'p16-r25' marginally (among 0.2\% difference) leads in WBC. 
Along with WSI and WSSS tasks at Table \ref{tab:table_wsi} and Table \ref{tab:table_WSSS_ablation}, a general trend is observed: the 'base' performs the highest, and the curriculum empowered 'p16-rd' experiments outperform the fixed design of 'p16-r25'.
Overall, the dynamic patch approach in PuzzleTuning amplifies multi-scale understanding. Through aggregating complexity, the task difficulty gradually increases, and such curriculum learning further enriches the puzzle understanding.
\section{Conclusion}
\label{sec:conclusion}

In conclusion, PuzzleTuning is proposed as a novel SSL pre-training strategy tailored for pathological image analysis. Specifically, we devise an explicit multiple puzzle restoring task focusing on grouping, junctions, and semantic alignment relationships. Regulated by curriculum learning design, the prompt tokens are leveraged to explicitly hone domain-bridging knowledge, seaming general vision with specific downstream pathological tasks. 
Extensive experiments are conducted with large-scale SSL pretraining and various downstream tasks with multiple datasets. The significant improvements in various applications prove its effectiveness and underscore the deliberate explicit design intentions.
\section*{Acknowledgements}
\label{sec:acknowledgements}

This work was partially supported by the National Natural Science Foundation of China (No. 62271023), the Beijing Natural Science Foundation (No. 7242269), the National Key Research and Development Program of China (No. 2023YFF0715400), and the Fundamental Research Funds for Central Universities. This work was also partially supported by the Biomedical Research Council of A*STAR (Agency for Science, Technology and Research), Singapore.

Several collaborators have contributed to this work, specifically, we appreciate the early-stage support from our collaborator Chunhui Li from the School of Artificial Intelligence, Nanjing University. We appreciate the dedicated suggestions from Dr. Liu Wei, and Dr. Malay Singh as well as other members from the CVPD lab from Bioinformatics Institute (BII), A*STAR.


 
%

\bibliographystyle{IEEEtran}
\bibliography{IEEEabrv,main}

\end{document}